\documentstyle[12pt,aaspp4]{article}

\newcommand{\etal}{{\it et al.}\ }

\lefthead{von Hippel \& Sarajedini}
\righthead{NGC 188}

\begin{document}

\title{WIYN Open Cluster Study 1: Deep Photometry of NGC 188}

\author{Ted von Hippel}
\affil{Department of Astronomy, University of Wisconsin, Madison WI 53706 \\
email: ted@noao.edu}
\authoraddr{National Optical Astronomy Observatories, 950 N. Cherry Av.,
Tucson, AZ 85719}

\author{Ata Sarajedini}
\affil{Department of Physics \& Astronomy, San Francisco State University, 
1600 Holloway Ave, San Francisco, CA 94132 \\
email: ata@stars.sfsu.edu}

\begin{abstract}

We have employed precise and carefully calibrated V- and I-band photometry
of NGC 188 at WIYN to explore the cluster luminosity function and study
the cluster white dwarfs.  Our photometry is offset by V = $0.052$
(fainter) from Sandage (1962) and Eggen \& Sandage (1969).  All published
photometry for the past three decades have been based on these two
calibrations, which are in error by $0.05 \pm 0.01$.  We employ the
Pinsonneault \etal (1998) fiducial open cluster main sequence to derive a
distance modulus of $11.43 \pm 0.08$ and E(B$-$V) = $0.09$, with the
largest single source of error caused by uncertainty in the cluster
metallicity.  We report observations that are $\geq 50$\% complete along
the main sequence to V = $24.6$.  We find that the NGC 188 central-field
LF peaks at M$_I \approx 3$ to $4$.  This is unlike the solar neighborhood
LF and unlike the LFs of dynamically unevolved portions of open and
globular clusters, all of which typically rise continuously until M$_I
\approx 9.5$.  Although we find that $\geq 50$\% of the unresolved objects
in this cluster are multiple systems with mass ratios $\geq 0.3$, their
presence cannot account for the shape of the NGC 188 LF.  For theoretical
reasons (Terlevich 1987; Vesperini \& Heggie 1997) having to do with the
long-term survivability of NGC 188 we believe the cluster is highly
dynamically evolved and that the low luminosity stars missing from the
central cluster LF are either in the cluster outskirts, or have left the
cluster altogether.  We identify nine candidate white dwarfs (WDs) in NGC
188, of which we expect at least three, and perhaps six, are bona fide
cluster WDs.  The luminosities of the faintest likely WD indicates an age
(Bergeron, Wesemael, \& Beauchamp 1995) of $1.14 \pm 0.09$ Gyrs, where the
error in age includes the cluster distance uncertainty and we assume the
WD has a Hydrogen atmosphere.  This age is a lower limit to the cluster
age and observations probing to V = $27$ or $28$ will be necessary to find
the faintest cluster WDs and independently determine the cluster age.
While our lower age limit is not surprising for this $\approx 6$ Gyr old
cluster, our result demonstrates the value of the WD age technique with
its very low internal errors.

\end{abstract}


\section{Introduction}

NGC 188 has been the subject of frequent and intensive investigation
primarily due to its great age.  Many early studies (e.g., Sandage 1962;
Eggen \& Sandage 1969; Spinrad \& Taylor 1969; Spinrad \etal 1970) were
motivated largely by the belief that NGC 188 was approximately $10$ Gyrs
old and therefore perhaps the oldest observable open cluster.  Subsequent
work has shown that the early age estimates were too large, with more
modern values being typically $6$ or $7$ Gyrs (e.g., Twarog \&
Anthony-Twarog 1989; Caputo \etal 1990; Demarque, Guenther, \& Green 1992;
Carraro \etal 1994; Dinescu \etal 1995).  More modern work has also shown
that other clusters such as NGC 6791 and Berkeley 17 (see Phelps, Janes,
\& Montgomery 1994) are older than NGC 188.  While NGC 188 has lost its
pivotal role in constraining the age of the Galactic disk, it has still
remained critically important for age studies since it is both nearer and
less obscured than older clusters.  Its metallicity is also typical of
the solar neighborhood with representative modern quotes of [Fe/H] = $0.02
\pm 0.11$ (Caputo \etal 1990), $-0.12 \pm 0.16$ (Hobbs, Thorburn, \&
Rodriguez-Bell 1990), and $-0.02$ (Twarog, Ashman, \& Anthony-Twarog
1997).  Interest in NGC 188 has not been limited to its age and role in
Galactic disk evolution, of course.  A number of studies have looked at
the cluster's wide giant branch (e.g., Twarog 1978; Norris \& Smith 1985),
blue stragglers (e.g., Leonard \& Linnell 1992; Dinescu \etal 1996), and
numerous W Ursa Majoris type and other contact binaries (e.g., van 't Veer
1984; Baliunas \& Guinan 1985; Moss 1985; Kaluzny \& Shara 1987; Kaluzny
1990).  The great age of NGC 188 also means that cluster Lithium
observations (e.g., Hobbs \& Pilachowski 1988) can
constrain stellar evolution and possibly cosmology.  Extensive proper
motion observations (Upgren, Mesrobian, \& Keridge 1972; Dinescu \etal
1996) and radial velocity work (e.g., Scott, Friel, \& Janes 1995;
Mathieu \& Dolan 1998)
have also sought to help with cluster membership issues and the
nature of the cluster's orbit in the Galaxy (e.g., Keenan, Innanen, \& House
1973; Carraro \& Chiosi 1994).

All of the above-quoted studies have necessarily been based on cluster
members brighter than V $\approx 20$, and often brighter than V $\approx
15$.  Our purposes in this paper were to add to our knowledge of NGC 188
based on the modern techniques of deep CCD photometry coupled with a large
telescope.  Specifically we wanted to address the cluster luminosity
function, which is the result of both the initial mass function (IMF) and the
cluster dynamical evolution.  Our goal was to observe to $\sim 10$ mags
fainter than the cluster turn-off, or to an equivalent mass of $\sim 0.2
M_{\sun}$.  These observations could then be compared with other deep
photometry of open and globular clusters, obtained primarily with the
{\it Hubble Space Telescope} (HST).  We were able to achieve this goal in
a central cluster field and discuss the ramifications for the dynamical
evolution of NGC 188 below.  We also sought to find the faintest possible
cluster white dwarfs.  Recent HST observations (von Hippel, Gilmore, \&
Jones 1995; Richer \etal 1997) have demonstrated that cluster white dwarfs
are a powerful and independent means of determining the ages of star
clusters.  In addition, while completing this paper we came across a
comparative study of open cluster distances based on {\it Hipparcos} (ESA
1997) parallax measurements and the main-sequence fitting technique
(Pinsonneault \etal 1998) which allows us to independently rederive the
distance to NGC 188.  An independent distance to NGC 188 is important as
there has been considerable difficulty and uncertainty in untangling the
cluster's distance, reddening, and metallicity (see Twarog \&
Anthony-Twarog 1989).  Finally, we do not fit isochrones as the turn-off
and giant branch are not well delineated in our data.  Isochrone fitting
will be the subject of a future paper 
which will combine deep WIYN data with wide-field photometry from the Kitt
Peak 0.9m telescope.

Our paper is also the first in what we hope will be a long and useful
series of coordinated papers on open clusters relying heavily on
WIYN\footnote{The WIYN Observatory is a joint facility of the University
of Wisconsin-Madison, Indiana University, Yale University, and the
National Optical Astronomy Observatories.} data.  With its one degree
diameter field-of-view feeding a $100$ fiber multi-object spectrograph,
its high quality imaging, and sizable aperture WIYN is an ideal tool for
open cluster work.  To this end a group of approximately a dozen
researchers from the WIYN institutions and their collaborators have formed
what we call the WIYN Open Cluster Study (WOCS).  Our collaboration will
study approximately $10$ open clusters spanning the widest possible range
of age and metallicity in great detail over the next decade, measuring Fe,
C, O, Li, and hopefully N abundances, precise radial velocities, proper
motions, and broad-band colors.  While the observations are to take place
primarily at WIYN, we will also supplement and support our WIYN
observations with other telescopes.  We will present our results and data
on-line (see http://www.astro.wisc.edu/) as they are published.

\section{Data Reduction}

We observed NGC 188 at the WIYN 3.5m telescope on Kitt Peak through the
Harris V-band and Mould interference I-band filters (Massey \etal 1987)
over a 14 month period during runs ranging from a few hours to a few
nights in duration.  Table 1 lists the details of the observations
including the four night period August 12-15, 1996, when low transparency
and poor seeing in the end made the observations too poor for our use.
The first column of Table 1 lists the date(s) of the runs, the second
column lists the sky conditions with ``P'' indicating a photometric night
and ``NP'' indicating a non-photometric night, the third and fourth
columns list the exposure times in hours in the V- and I-bands,
respectively, the fifth column lists the range of seeing for the run
measured as the FWHM of a Gaussian fit, and the sixth column provides
notes to the runs.  The table includes 11 nights, a few of which were
partially allocated to this program.  Of the cumulative 10 nights devoted
to this project, due to frequent bad weather and some problems with the
CCD (discussed below), the final tally of useful observations was 5.75
hours of V-band exposure and 6.5 hours of I-band exposure.  The image
quality was also generally poorer than the median recorded at WIYN
($\sim0.8${\arcsec} FWHM), ranging from $0.7$ to $1.6${\arcsec} FWHM.
Nonetheless, $\sim6$ hours of exposure per filter with a $3.5$ meter
telescope allows us to address a number of the goals as set out in the
introduction and to present observations of this cluster to V $\approx
25$, five magnitudes fainter than previous work.

The CCD detector currently in use at WIYN, ``S2KB'', is a $2048^2$ STIS
CCD with $21$ micron (=$0.2${\arcsec}) pixels and a field of view of $6.8
\times 6.8$ arc minutes.  Observations were obtained near the center of
NGC 188, at RA = $0^{h}48^{m}26^{s}$, Dec =
$+85{\arcdeg}15{\arcmin}21{\arcsec}$ (J2000), corresponding to Galactic
coordinates {\it l, b} = $122.9{\arcdeg}$, $+22.4{\arcdeg}$.  Data reduction followed
standard procedures although a few extra steps for the particular
characteristics of the S2KB CCD were incorporated.  In particular, this
CCD has a time-dependent, two-dimensional bias structure that can range
from $\approx 4$ ADU to $\approx 20$ ADU.  The bias structure is typically
stable within a few ADU on any given night, however.  By fitting a
polynomial of order $\sim 4$ to the overscan regions of all calibration
and science frames and performing a residual bias subtraction, we were
able to remove the bias offset and structure typically to within $2$ ADU.
The largest bias uncertainty was in the low row, low column number region,
a region we avoided using for the standard stars.  For the deeper
broad-band (science) exposures the high sky meant that this bias
uncertainty was always much less than $1$\%.  Flat fielding was performed
using dome flats with typical pixel-to-pixel Poisson uncertainties of less
than $0.3$\% and illumination pattern uncertainties less than $1.0$\%.

All photometry over the past 30 years of NGC 188 with which we are
familiar has been calibrated against Sandage (1962) and Eggen \& Sandage
(1969).  Since photometric equipment and techniques have improved over the
past 30 years we decided to obtain our own photometric calibration at
WIYN.  Our early efforts were hampered by a (then unknown) severe
non-linearity in the S2KB CCD when this chip was exposed to background
counts of less than $\approx 200$ ADU (at an inverse gain setting of $2.8$
e-/ADU).  The non-linearity was background dependent and rose to $40$\%
for point sources on a zero count background.  We thus discarded our short
calibration exposures taken during July 24-27, 1995 (Table 1) and
reobserved NGC 188 on the photometric night of September 19, 1996 once the 
S2KB CCD had been
repaired.  On September 19, 1996 we also obtained $46$ V-band observations of
$25$ Landolt (1983, 1992) standard stars and $40$ I-band observations of
$26$ Landolt standard stars.  These observations were reduced in the same
manner as the data, although with the aperture corrections determined from
the standard stars rather than from the bright program stars.  We used the
routine of Harris, Fitzgerald, \& Reed (1981) to simultaneously fit a zero
point, an airmass term, and a first order color term for each of V and I.
The resulting photometric transformations were

\begin{equation}
   V = v + 24.877 - 0.169 \ X + 0.010 \ (V-I) \ and
\end{equation}

\begin{equation}
   I = i + 24.271 - 0.065 \ X + 0.031 \ (V-I)
\end{equation}

\noindent
where V and I are magnitudes in the standard system, v and i are
magnitudes in the instrumental system, and X is the airmass.

After applying the above calibration to our data we looked for a range of
parameters that might correlate with the residuals in the photometry.
Nothing we checked correlated with the photometric residuals except
universal time (UT).  Not only did we search for correlations using the
standard star data, but we also used many bright stars in the NGC 188
frames themselves ($15$ epochs) to look for the same correlations since
the entire night of September 19, 1996 was spent monitoring the same field
in NGC 188 and standard stars.  We found a highly significant, but
small-valued, correlation between universal time and the residuals in both
the NGC 188 stars and the standard stars.  This correlation is shown in
Figure 1.  Since the NGC 188 observations cover a greater range in
universal time, they better determined the slope of the UT versus residual
V and I magnitude relations.  The NGC 188 residual slope was imposed on
the standard star data.  Figure 1 demonstrates that the NGC 188 and
standard star residual slopes were very similar, however.  The best fit
correlations were

\begin{equation}
   cor(UT,V) = 0.1275 - 0.0125 \ UT(V) \ and
\end{equation}

\begin{equation}
   cor(UT,I) = 0.1020 - 0.010 \ UT(I)
\end{equation}

\noindent
where cor(UT,V) and cor(UT,I) are the corrections to apply as a function
of universal time in V and I, respectively, and UT(V) and UT(I) are the
values of universal time during the observations in V and I,
respectively.  The additive constants in the above two equations zero
point the photometry to UT = $10.20$.  The NGC 188 data were offset to the
same zero point as the standard stars.

We interpret the temporal dependence of the photometry on a slow and
monotonic change in the atmospheric transparency.  The final result had
internal residuals of $0.004$ mag for the V-band and $0.009$ mag for the
I-band.  The color terms of $0.010$ and $0.031$ for V and I, respectively,
were consistent with the color terms measured by our group during
telescope commissioning and seen subsequently by other observers.  Both
the small residuals and the small color terms demonstrate that the WIYN
Harris V and Mould interference I filters are on the Landolt
Johnson-Kron-Cousins photometric system (see Landolt 1992).  The
atmospheric extinction terms are within the normal range of extinction
terms seen at Kitt Peak.

In Figure 2 we compare our V-band data with that of Sandage (1962) and
Eggen \& Sandage (1969) as a function of apparent V-band magnitude and as
a function of V$-$I color.  We find a mean offset of V(Eggen \& Sandage) $-$
V(this work) = $-0.052$ mag with a
standard deviation of $0.063$ mag.  For the $76$ objects in common the
offset is highly significant.  (When we compare our photometry to the $23$
stars in common with Caputo \etal (1990) we find a mean offset of $-0.059$
and a standard deviation of $0.022$.  The similarity is expected since the
Caputo \etal photometry was calibrated against that of Eggen \& Sandage.)
These data show no convincing trend between the difference of our
photometry and that of Eggen \& Sandage with either magnitude (Figure 2a)
or color (Figure 2b).  We attribute the systematic photometric difference
between our work and that of Eggen \& Sandage to possible differences in
the photoelectric versus CCD passband and calibrations, as well as the
difficulty of doing photometry at an airmass of $\geq 1.7$.  As a further
check we compared V-band magnitudes for stars in common between our WIYN
data and our (as yet) unpublished KPNO 0.9m data, employing a calibration
kindly provided by E.M. Green (private communication).  We found that these two
independent calibrations agree to within $0.01$ mag.  We believe the
modern filter prescription, our careful tying of the WIYN photometry to
the Landolt system, and the modern ability to place multiple standard
stars on the same CCD field, allowed us to determine a more reliable
photometric zero point than that of Sandage (1962) and Eggen \& Sandage
(1969).

For each of the observing runs (Table 1) we coadded the data within a
given run using IRAF (Tody 1993) routines to register, scale, and offset
the frames.  Scaling and offsetting the frames before coaddition is
necessary since the sky transparency and background change as a function
of airmass even under photometric conditions.  The resulting V- and I-band
frames for each run were then fit with a quadratically variable point
spread function (PSF) using the multiple-PSF fitting ALLFRAME reduction
package (Stetson 1994, see below).  We employ the PSF-fitting photometry
rather than aperture photometry due to its optimum signal-to-noise
weighting of the data, and not because of crowding.  These images contain
$\lesssim 10^4$ objects spread over $4 \times 10^6$ pixels, and so only a
small percentage of the detected objects are blended with other objects.

For each combined image, we began by constructing a high signal-to-noise
PSF using $50$ to $100$ bright uncrowded stars.  We then established a
spatial transformation between each combined image and a reference image
using the DAOMASTER software kindly provided by P. Stetson.  We use this
transformation to further combine all of the combined images to form one
master frame to be used in defining a list of coordinates for all detected
image profiles.  The SExtractor (Bertin \& Arnouts 1996, see below) source
finding routine was used to expedite this process.  This master coordinate
list was input into ALLFRAME along with the PSFs of each frame and the
spatial transformation relation.  The output of ALLFRAME consists of
instrumental photometry for each detected profile on each frame as well as
an image from which all of the profiles have been subtracted.  We then
combined all of these subtracted frames and used SExtractor to augment the
original list of image profiles.  This new master list of coordinates was
again fed into ALLFRAME to yield the final list of instrumental
magnitudes.

Because of the slight warpage of the S2KB CCD the quadratically varying
PSF did not fully account for the spatial variation of the PSF.  To
correct for this, we selected between $200$ and $300$ bright stars per
epoch.  After all of the other stars were subtracted from the frames, we
performed large-aperture (radius = $25$ pixels) photometry for all of
these stars.  The difference between the PSF magnitude and the aperture
magnitude [cor(R)]  was then plotted as a function of the distance from
the lower-left corner of the frame (R). The only frame which exhibited a
large variation ($\pm0.05$ mag) was the I frame taken on February 4,
1996.  The others showed a very small $\pm0.005$ mag variation.  In any
case, we fitted polynomials of the form

\begin{equation}
   cor(R) = a_0 + a_1 \ R + a_2 \ R^2
\end{equation}

\noindent
to the data using an iterative $2\sigma$ rejection technique. These
equations were applied to the instrumental magnitudes.

The photometry for each run was then merged and put on the Landolt
standard system based on the bright, but unsaturated, stars that
overlapped with the September 19, 1996 data set.  The number of such stars
ranged from $41$ to $119$ per run and the residuals in the offset range
from $0.015$ to $0.03$ mag.

Since a large number of the detected faint objects are not stars, but
rather background galaxies, we also required a method to morphologically
reject non-stellar objects.  After careful comparison of visual and
automated morphological rejection techniques we chose to use SExtractor
(Bertin \& Arnouts 1996).  SExtractor has the advantage of determining sky
locally and uses verified neural network techniques to perform the star
galaxy classification.  Figure 3 shows the results of the SExtractor
morphological classification versus I-band magnitude.  The ``Stellarity
Index'' ranges from $0$ (galaxies) to $1$ (stars).  Although SExtractor's
neural network classifier is not strictly a Bayesian classifier, the
Stellarity Index values are approximately the probabilities that the
object is a point source.  The large number of definite stars in Figure 3
demonstrates that a cluster is present in this field, though there is
significant contamination, especially at the faintest magnitudes.  The
large number of classifications near $0.5$ at the faintest magnitudes
demonstrates the common sense notion that at limiting signal-to-noise
classification breaks down.  SExtractor understands this and yields a
sensible classification of $\sim0.5$, i.e., unsure.  Note also that
saturated stars have an associated probability of being stellar of
somewhat less than $0.9$.  This too is sensible behavior for SExtractor
given that these stars have flat-topped, broader PSFs.

We select only those objects determined by SExtractor to have a high
($\geq 0.95$, marked by the dashed line in Figure 3, i.e., $\gtrsim 95$\%)
probability\footnote{We performed careful visual examination of the images
and found two galaxies and one location on a diffraction spike for which
SExtractor reported stellarity $\geq 0.95$.  The misclassification is
consistent with the probablistic nature of the classification and the
stellarity $\geq 0.95$ cut-off.} of being stars and present their
calibrated color magnitude diagram (CMD) in Figure 4.  Note that the
requirement of a good morphological classification, which is somewhat
different than simply signal-to-noise, imposes the faint magnitude limit
in Figure 4.  Before moving on to a detailed discussion of the NGC 188 CMD
we pause to point out that a strong binary sequence is evident in this
cluster to the limit of the data and that the number of cluster stars
drops markedly fainter than V $\approx 18$.  The turn-off for this cluster
is at V $\approx 15$ and the faintest stars at V = $25$ have masses
$\lesssim 0.2 M_{\sun}$ based on the empirical mass-luminosity relations
of Henry \& McCarthy (1993).  Individual error bars are not plotted for
clarity but typical photometric errors at each integer V-band magnitude
starting at V = $19$ are presented down the right-hand side of Figure 4.
These errors are internal errors only.

We estimate the total systematic error of our photometry as due to the
errors in the standard star calibration and the errors in transforming the
data taken under non-photometric conditions onto the single calibrated
night.  The external accuracy of the standard star calibration should be
excellent since the residuals about the final calibration were only
$0.004$ in V and $0.009$ in I.  Furthermore, dropping individual standard
stars from the solution affected the transformation coefficients by always
$\leq 0.003$ mag.  Note, however, that the reddest Landolt standard we
used had V$-$I = $2.08$, whereas our data continue to V$-$I = $3$, and
beyond.  The color term for the lower main sequence is thus an
extrapolation and is likely to be somewhat less accurate than the rest of
the photometry.  We estimate that our transformations blueward of V$-$I =
$2$ to the Landolt system are accurate to $\leq 0.01$ mag.  We further
estimate that the procedures of calibrating subsequent photometry to the
one calibrated night are accurate to $0.01$ to $0.03$ mag, depending on
the epoch.  Multiple observations for most stars will reduce the
epoch-to-epoch variations, but we will use $0.03$ mag as our error for
this bootstrapping calibration step.  The overall external systematic in
the photometry blueward of V$-$I = $2$ should be $\leq 0.032$.

\section{Discussion}

\subsection{A New Distance Determination}

Past distance determinations (see Twarog \& Anthony-Twarog 1989 for an
extensive discussion) to NGC 188 have been based on the morphology of the
turn-off, subgiant, and red giant regions fit by stellar isochrones.
These fits simultaneously determine the cluster distance, reddening, and
age, while explicitly adopting a metallicity value and implicitly
depending on a host of parameters in the stellar evolution models.
Perhaps the most up-to-date and reliable value for the distance and
reddening of NGC 188 is m$-$M = $11.35$ and E(B$-$V) = $0.09$ (Twarog
\etal 1997).  \footnote{We rely on the reddening relations of Cardelli,
Clayton, \& Mathis (1989) to convert E(B$-$V) to E(V$-$I), see below.}
These values update the results of Twarog \& Anthony-Twarog (1989), who
found m$-$M = $11.50$ and E(B$-$V) = $0.12$, which since 1989 have been
the canonical cluster values.  While Twarog \& Anthony-Twarog (1989) and
Twarog \etal (1997) have considered other constraints on the cluster
distance and reddening besides fits to stellar evolutionary models, the
precise distance, and especially reddening, still remain uncertain.  It is
our contention that the distance modulus is uncertain to $\sim 0.10$ mag
and the reddening to $\sim 0.03$ mag.  In addition, since our photometry
is offset by V = $0.052$ (fainter) from Sandage (1962) and Eggen \& Sandage
(1969), it is necessarily offset by the same value from the photometric
distance determinations of all other studies since they were calibrated
against Sandage (1962) and Eggen \& Sandage (1969).  For all of these
reasons, and since we can provide a distance determination independent of
previous techniques, we rederive the distance to NGC 188, below.

Our data precisely determine the main sequence locus from the turn-off
near V = $15$ to the lower main sequence at least as faint as V = $21$
(where V$-$I $\approx 2$, the interpolation limitation of the Landolt
standards).  We would like to compare our photometry to the precise
trigonometric parallaxes of solar metallicity dwarf stars made by the
{\it Hipparcos} (ESA 1997) mission.  Since the {\it Hipparcos} satellite
did not make photometric measurements in the Johnson-Kron-Cousins system
there is no directly determined fiducial main sequence for our filters.
Fortunately, Pinsonneault \etal (1998) have just completed a detailed
study of {\it Hipparcos} parallaxes to individual cluster members in five
nearby open clusters ($\alpha$ Per, Coma Ber, Hyades, Pleiades, and
Praesepe) and they have derived just such a fiducial main sequence for
solar metallicity stars in both B$-$V and V$-$I.  Although Pinsonneault
\etal find a systematic problem with the Pleiades parallaxes, they argue
that the other four open clusters are consistent with the same fiducial
sequence within very stringent limits.  They conclude that with good data
one can derive the distance to a near solar metallicity open cluster using
the main-sequence fitting technique to an accuracy in the distance modulus
of $0.05$ mag.  Since we have no independent means of determining the
reddening to NGC 188 we derive the distance based on the two recent and
most commonly quoted values, E(B$-$V) = $0.09$ and $0.12$.  In Figures 5a
and 5b we present our photometry in the region of the Pinsonneault \etal
(1998) fiducial solar metallicity main sequence (solid lines) for our best
fit distance of m$-$M = $11.43$ for E(B$-$V) = $0.09$ and m$-$M = $11.67$
for E(B$-$V) = $0.12$, respectively.  The one sigma error bars are plotted
for the data and fiducial sequences offset by $\pm 0.05$ mags are
presented as dashed lines.

We derived the cluster distance moduli by first selecting what appeared to
be single-star cluster main sequence members within the range $0.8 \leq$
V$-$I $\leq 1.0$, i.e., unevolved main sequence stars within the
Pinsonneault \etal color calibration range ((V$-$I)$_o \leq 0.9$).  These
selected stars are indicated in Figure 5 by overplotted circle symbols.
We then slid the fiducial sequence vertically in $0.01$ mag increments
until the mean offset of the selected stars went to zero.  Following this
procedure we determined initial cluster distance moduli of m$-$M = $11.41$
for E(B$-$V) = $0.09$ and m$-$M = $11.63$ for E(B$-$V) = $0.12$.  Since
photometric binaries create a sequence of objects above and to the right
of the main sequence we then studied the distribution of our selected
stars about the fiducial sequence for the initial cluster distance
moduli.  These distributions are plotted in Figures 6a \& 6b for the
E(B$-$V) = $0.09$ and E(B$-$V) = $0.12$ cases, respectively.  One can
clearly see in both figures that the stars do not comprise a normal
distribution about the initial cluster distance moduli.  Rather, a peak at
negative offset values is present, as is a tail to positive offset
values.  This is precisely what we expect from the inclusion of unresolved
binaries.  We thus improve upon our initial distance estimate by including
an offset of $0.02$ and $0.04$ mag to the peak in the single star sequence
for the two E(B$-$V) = $0.09$ and $0.12$ cases, respectively.  We also
performed these same tests with our unpublished KPNO 0.9m photometry and
found  m$-$M = $11.43$ for E(B$-$V) = $0.09$ and m$-$M = $11.58$ for
E(B$-$V) = $0.12$.  The consistency between the WIYN and KPNO 0.9m
E(B$-$V) = $0.09$ results is encouraging, as is the fact they they agree
with the Twarog \etal (1997) results (recall the $0.05$ mag offset in the
photometric zero point).  The $0.05$ mag difference between the WIYN and
KPNO 0.9m E(B$-$V) = $0.12$ results does not disfavor these results, as
this difference is within the quoted errors of the technique, and could
anyway result from a departure from solar metallicity.  We believe,
however, that m$-$M = $11.63$ is too high given the extensive past work on
the cluster distance, while at the same time the lower reddening seems to
be preferred by Twarog \etal (1997).  We thus rely on m$-$M = $11.43$ and
E(B$-$V) = $0.09$ for the remainder of this paper.

The quality of our photometry is good enough that we adopt the $0.05$ mag
precision for this distance technique quoted by Pinsonneault \etal
(1998).  One external error to consider is uncertainty in the ratio of
E(V$-$I) to E(B$-$V).  We rely on equations 3a \& 3b of Cardelli, Clayton,
\& Mathis (1989) to derive a value of $1.34$ for the central wavelength
($8220$ \AA) of our I-band filter.  This may be mildly inconsistent with
Pinsonneault \etal who use E(V$-$I) = $1.25$ E(B$-$V), though we cannot be
sure as we do not know the central wavelength(s) for the I-band photometry
they have collected.  We estimate this source of uncertainty at a given
E(B$-$V) value to be of order only $0.01$.  Another source of external
error is the external accuracy of our photometry.  We argued above that
this should be $\leq 0.032$.  Yet another source of external error is
uncertainty in the metallicity of NGC 188.  If we take the mean and range
of the above-quoted three metallicity values and employ small sample
statistics (Keeping 1962) we estimate the metallicity of NGC 188 to be
$-0.04 \pm 0.05$ (standard error in the mean).  Alternatively, we could 
rely solely on 
the best recent spectroscopic metallicity estimate, which is 
$-0.12 \pm 0.16$ (Hobbs, Thorburn, \& Rodriguez-Bell 1990).
Both means of determining the cluster metallicity provide values that are 
solar to within
the errors and so we will not make a metallicity correction.  We note,
however, that there is still necessarily uncertainty in the metallicity
for NGC 188.  If we take twice the standard error listed above (i.e., $\pm
0.10$) as representative of this uncertainty, this leads to a systematic
uncertainty in the distance of $0.06$ mag (Pinsonneault \etal 1998).
Combining all of these sources of error in quadrature we estimate that the
distance modulus to NGC 188 is $11.43 \pm 0.08$ for E(B$-$V) = $0.09$,
with the largest single source of error being the uncertainty in the cluster
metallicity.

\subsection{Removing Field Star Contamination}

Although the cluster main sequence is obvious in Figure 4, especially at
the brighter magnitudes, a moderate number of field stars are also
apparent throughout the CMD.  Most of the field stars are fainter than the
cluster main sequence, which is as expected since the volume surveyed
increases with distance.  The vast majority of the field star
contamination is due to main sequence stars from the Galactic disk, though
a few thick disk and halo main sequence stars most likely also are
present.  Since some Galactic field stars overlap the position of the main
sequence a procedure is required to remove, at least statistically, the
Galactic field star component.  The standard procedure of statistical
field star removal is to observe a comparably deep field outside the
cluster at the same Galactic latitude.  Because of the difficulties we had
in obtaining these data we were unable to obtain a comparison field.
Nonetheless, at least for our purposes (extracting the cluster main
sequence and white dwarfs), we can adequately subtract field stars with
the aid of Galaxy models.  Figure 7 shows the CMD for the Galaxy model of
Reid \& Majewski (1993) for the location, sky coverage, and reddening of
our NGC 188 field.  This model has been tested against north Galactic pole
number counts and color distributions (Reid \& Majewski 1993) and against
two deep, lower latitude fields (Reid \etal 1996).  For our field the
number and distribution of model field stars also appears very similar to
the observed distribution of field stars in Figure 4.  The only location
in the CMD where there appears to be some disagreement is for stars
considerably redder than the NGC 188 main sequence.  These objects are
unlikely to be cluster members, yet no comparable population is seen in
the Reid \& Majewski Galaxy model.  Despite these missing model stars, for
our purposes the Reid \& Majewski model does an excellent job of mimicking
the observed cluster field stars.  As an additional check we also
calculated the same Galaxy field CMD with the Gilmore, Reid, \& Hewitt
(1985; see also Gilmore, Wyse, \& Kuijken 1989) Galaxy model.  Although
the Gilmore \etal model provides B$-$V colors, rather than V$-$I colors,
we carefully mapped B$-$V to V$-$I for main sequence stars and found the
two model CMDs to give essentially identical results for our field.

Our procedure for determining the cluster main sequence luminosity
function (LF) was to first isolate the region of the observed CMD that
contained the cluster main sequence, then to isolate the same region in
the Reid \& Majewski model CMD.  We then subtracted the model population
from the observed population in $0.5$ mag bins.  This procedure should be
insensitive to precisely how we isolate the main sequence in the CMD as
long as we do not isolate too narrow a region, and thereby miss some of
the observed main sequence stars.  Isolating an overly large portion of
the CMD will cause us to count more real field stars and more model field
stars, but the difference between the two is still the number of NGC 188
main sequence stars.  We thus chose to consider main sequence cluster
members to reside throughout a somewhat larger area than one might draw by
eye, bounded on the blue side by a line starting at V, V$-$I = $15$, $0.2$ and
bounded on the red side by a line with the same slope shifted by V$-$I =
$0.7$ redward.  The turn-off region and stars brighter than V = $15$ were
included in their entirety, although two or three of these stars may be
non-members.  Before continuing on to present the cluster luminosity
function, we take up the important issue of observational completeness.

\subsection{Determining Completeness}

In order to interpret the cluster luminosity function in an
astrophysically useful manner, we must understand and account for the
photometric completeness of the data.  This means that we need to
determine the efficiency of our star detection and measurement procedure
at each magnitude.  To accomplish this we adopted the standard approach of
artificial star experiments.  The general idea is to construct a set of
artificial stars with known magnitudes and positions, place these on the
cluster frames, and reduce the resulting images in precisely the same
manner as the original data.

We began by constructing a fiducial of the NGC 188 main sequence extending
from V = $15$ to V = $26$. We selected stars on the fiducial sequence in
half-magnitude bins between $15 <$ V $< 20$ and quarter-magnitude bins
between $20 <$ V $< 26$.  Each bin in the former magnitude range contains
$38$ artificial stars and each bin in the latter contains $24$ such stars
totaling to $956$ artificial stars. The positions of these stars were
randomly generated but were constrained not to fall within $1.5$ PSF radii
of each other.  These stars were then placed on each of the frames using
the ADDSTAR routine in DAOPHOT II (Stetson 1994), which incorporates the
appropriate amount of noise. The resulting frames were reduced using
exactly the same procedure as the original ones. After comparing the input
into the artificial star tests with the output, we derive the relative
completeness curves shown in Figures 8a and 8b.  We find that the LFs are
$\geq 50$\% complete at V = $24.6$ (I = $21.5$).

\subsection{The Main Sequence Luminosity Function}

We chose to study and compare the NGC 188 LF with the LFs of other stellar
populations, rather than convert these LFs to mass functions, because of
the well-known sensitivity of mass functions to the exact shape of the
uncertain mass-luminosity relation.  The data presented in Figure 4 show
the main sequence of NGC 188 extending to V $\approx 25$ where it becomes
difficult to distinguish from the contaminating Galactic field stars.  It is
clear, however, that the luminosity function drops markedly for magnitudes
fainter than V $\approx 18$, and that an extensive binary sequence exists
to the limit of the data.  The cluster and model field star LFs, as
discussed above, are presented in Figures 8a \& 8b (V and I,
respectively).  The model Galaxy contribution increases slowly with
magnitude but until the observations begin to suffer from incompleteness
the model field contribution is always small compared with the observed
number counts.  The field star correction is thus modest and reasonable
errors in the Reid \& Majewski model should not affect the overall
results.  Figures 8c \& 8d present the completeness-corrected differences 
between the LFs of
Figures 8a \& 8b, respectively, with the x-axes now in absolute rather
than apparent magnitudes.  The V- and I-band LFs both peak at M$_V$ and
M$_I$ in the range of $3$ to $4$, and are plotted only where completeness
is $\geq 50$\%.  These LFs are unlike the solar neighborhood LF and unlike
the LFs from a broad range of open and globular clusters chosen such that
mass segregation and other dynamical processes are unimportant.  A number
of such representative LFs ($\omega$ Cen from Elson \etal 1995; 47 Tuc
from De Marchi \& Paresce 1995b; M 15 from De Marchi \& Paresce 1995a; NGC
6397 from Paresce, De Marchi, \& Romaniello 1995; NGC 2420 and NGC 2477
from von Hippel \etal 1996; the solar neighborhood from Kroupa, Tout, \&
Gilmore 1993) are plotted in Figure 8d where it can be seen that they
typically rise continuously until M$_I \approx 9.5$.

The above analysis presents the NGC 188 LF under the assumption that all
main sequence objects are single stars.  We can tell from a casual glance
at Figure 4, however, that for NGC 188 this is incorrect; the cluster has
a large population of binary star systems.  Open clusters typically have a
large fraction of binaries, while globular clusters typically have far
fewer ($\lesssim 10$\%, e.g., Richer \etal 1997, and references therein).  
The general effect of counting multiple star
systems as single stars in a LF is to overestimate the brightness of many
stars and to underestimate the contribution of low mass members.  While
this is undoubtedly causing the observed NGC 188 LF to be unrepresentative
of its underlying population, this problem also occurs to a similar degree
in the observed LFs for NGC 2420 and NGC 2477.  These open cluster LFs
look nothing like the NGC 188 LF, so although binaries may play a role in
changing the LF shape they are not responsible for the great differences
seen between LF shapes in Figure 8d.

While the statistical subtraction technique employed above does not allow
us to identify individual single and binary cluster members, it does allow
us to estimate the fractional contribution of binary stars.  The
distribution of objects within $0.2$ mag of the fiducial main sequence has
already been plotted in Figure 6 within the limited color range of $0.8
\leq$ V$-$I $\leq 1.0$, corresponding to the range $16.3 \leq$ V $\leq
17.3$.  This region appears to the eye to have approximately the same
binary contribution as fainter portions of the main sequence, and the
greater number of stars in this region make any fractional determination
more robust than at fainter magnitudes.  Examining Figure 6a we find $18$
($21$) stars in the leftmost three (four) bins, while the remainder of the
bins have $25$ ($22$) stars.  These numbers are equal within their
uncertainty.  Since the number gradient of field stars across the main
sequence is small, we assume the background field contributes equally to
the identified single and binary distributions.  We find that $\sim50$\%
of main sequence stars within $0.2$ mag of the main sequence are single
stars or have low enough mass ratios (q $\leq 0.3$, see Pols \& Marinus
1994) to appear single, and that the remaining $\sim50$\% are multiple
systems.  More multiple star systems are likely present at greater
luminosity separation from the main sequence, up to at least $0.75$ mag,
the expected offset caused by two equal mass binary members.  We assume
the binary fraction we measure is thus a lower limit for the $0.8
\leq$ V$-$I $\leq 1.0$ color region and we further assume that this region
of the main sequence is typical of the cluster.  We conclude that at least
$50$\% of the cluster objects are binaries.

Since the LFs presented here come from a single central cluster field it
is difficult to determine whether the deficit of low mass stars is due
simply to mass segregation (i.e., the low mass cluster members still belong
to the cluster but occupy a much larger volume), whether a significant
number of low mass members have been dynamically ejected, or whether the
cluster always had a deficit of low mass members.  This last possibility,
that NGC 188 was created with a deficit of low mass members, is unlikely,
however, since clusters with overly flat initial mass functions are
dynamically unstable (Terlevich 1987; Vesperini \& Heggie 1997).
Additionally, on observational grounds (see von Hippel \etal 1996; von
Hippel 1998) little variation is seen between cluster mass functions for
cases where one can isolate a sample which has suffered little dynamical
evolution.  The case for dynamical evolution for NGC 188 is strong.  Other
authors (e.g., Dinescu \etal 1996) have noted the advanced dynamical state
of the cluster, which is to be expected from its great age.  In fact, the
vast majority of open clusters do not survive to nearly the age of this
cluster.  McClure \& Twarog (1977) give a relaxation time for NGC 188 of
$6.4 \times 10^7$ years and Binney \& Tremaine (1987) note that cluster
evaporation times are typically $100$ times their relaxation time, i.e.,
essentially the age of the cluster.

Finally, we note that the expected number of background contaminating
galaxies (see below) in the faint main sequence portion of our cluster is
essentially zero, since very few galaxies are as red as V$-$I $\approx
3$.

\subsection{White Dwarfs}

We now turn to a study of the white dwarf region of the CMD (Figures 4 and
9) where we identify nine candidate WDs.  These nine candidate WDs are
identified by overplotted circles in Figure 9, which is the dereddened CMD with
the distance modulus of 11.43 subtracted.  (There are two WD candidates at
V$-$I, M$_V$ = $-0.04$, $12.06$ and $-0.03$, $12.08$ so only eight
distinct objects are visible in the CMD).  Also plotted in this figure is
the observed solar neighborhood WD sequence of Monet \etal (1992) and the
log(g) = $8$, Hydrogen atmosphere theoretical WD cooling sequence of
Bergeron \etal (1995).  The candidate WDs were identified based on their
photometric proximity to the Monet \etal and Bergeron \etal cooling
sequences.  We also visually inspected the location of the WD candidates
on the images.  Some of the candidates were too faint to visually ensure
that they were not instrumental artifacts.  We believe ALLFRAME can more
reliably compare signals from multiple V- and I-band images than we can do
visually, and so we believe all or most of our nine WD candidates are real
objects.  However, three of the WD candidates (5, 8, and 9, see Table 2)
have a somewhat lower likelihood of being cluster WDs either because of
their distance from the WD loci (candidate 5 is $2 \sigma$ distant in
V$-$I) or because they have similar V and I magnitudes to the more
abundant field stars (candidates 8 and 9).  The Reid \& Majewski and
Gilmore \etal Galaxy models also allow us to estimate the expected field
star contamination in the WD region.  The Reid \& Majewski model predicts
that two or perhaps three Galactic field WDs would lie in the same region
of the CMD as the cluster WDs.  The Gilmore \etal model similarly predicts
that one or two Galactic field WDs would lie in the same region of the CMD
as cluster WD candidates.  Despite this possible contamination, the nine
candidates WDs significantly exceed the expected number (two or three) of
non-cluster stars with similar colors.  A reasonable upper limit to the
number of candidate WDs that are true cluster WDs is six, with the three
most tentative candidate objects mentioned above being identified as the
expected interlopers predicted by the Reid \& Majewski and Gilmore \etal
Galaxy models.  A reasonable lower limit is probably three WDs.  In this
case the poorer candidates are field main sequence stars and another three
objects nearer the WD sequence are field WDs as predicted by the Galaxy
models.

In the above discussion we have ignored the possible contribution of
unresolved background galaxies.  Although the number of background
galaxies in any deep field is large, we have two methods to discriminate
against them.  Our first method is to rely on image morphology as many,
and perhaps most, galaxies are at least marginally resolved in our best
seeing ($0.7${\arcsec}) images.  In a study of deep galaxy counts Smail
\etal (1995, see their Figure 4) demonstrate that the stellar locus is
almost completely distinguishable from galaxies down to R = $24$ (V =
$24.5$ for the average background galaxy).  While their images were
obtained during both better seeing ($0.5${\arcsec} to $0.6${\arcsec}
versus $0.7${\arcsec}) and higher signal-to-noise than ours, it seems safe
to assume that our conservative morphological discrimination should remove
$90$ to $95$\% of the background galaxies.  Our second method is to rely
on color selection.  Fortunately for the white dwarf study, there are 
few faint galaxies with colors near the best WD candidates, V$-$I = $0.0$
to $0.4$.  Smail \etal (1995) find that $95$\% of galaxies at V $\approx
24.5$ lie within a few tenths of a magnitude color range bluer than V$-$I
$\approx 1.0$.  They do not report the fraction as blue as our WD
candidates but it seems safe to assume that number to be much less than
$1$\%.  Our two discriminatory tools should thus reduce the contaminating
background galaxy count by a factor of at least $10^3$.  The number of
background galaxies per magnitude reported by Smail \etal at V = $24.5$,
when scaled to the size of the WIYN S2KB field-of-view ($0.0129$ deg$^2$),
corresponds to $\sim 650$ objects.  The expected contamination in the WD
region is therefore $< 1$ object.  Of course, tracing the WD sequence to
fainter flux levels and redder colors would make this problem increasingly
difficult, since the number of background galaxies increases dramatically
at fainter magnitudes, and since the color discrimination would markedly
diminish.

Observed and derived properties for the nine WD candidates are listed in
Table 2.  The derived properties are based on the models of Bergeron \etal
(1995) under the assumption that each object is indeed a white dwarf.  For
each WD candidate Table 2 lists the candidate identification number in
column one, the dereddened V$-$I color in column two, the absolute V-band
magnitude in column three, the effective temperature in Kelvins in column
four, the WD mass in solar masses in column five, the absolute bolometric
magnitude in column six, and the logarithm of the age in years in column
seven.  Columns three, four, six, and seven also list errors in the
observed and derived quantities in parentheses.  The errors in the derived
quantities are based on propagating the observed photometric errors
through the Bergeron \etal model calibrations; as such they are internal
errors only.  No errors are listed for the derived quantity of mass since
formally the error is always $0.001$ solar masses.  Note that the mass
values are the result of using models with fixed log(g) = $8$, and not the
result of determining the expected mass for each WD based on the cluster
age and WD cooling ages.  Nonetheless, the derived WD masses are all nearly
$0.6$ solar masses, and these minor variations are unimportant for our
purposes, which is to comment generally on the white dwarf ages.  The
derived quantities are listed twice, based first on Hydrogen and then on
Helium atmosphere models, as indicated in column eight.  The vast majority
of spectroscopically observed WDs have Hydrogen atmospheres so these are
the numbers we expect to be relevant for NGC 188.  The Helium atmosphere
results are included to demonstrate the size of the effect of this basic
change in the WD parameters.

If all the objects listed in Table 2 are cluster WDs then NGC 188 must be
at least as old as its faintest WD.  The lower age limit would be $1.91
\pm 0.16$ Gyrs (H atmosphere result) or $2.58 \pm 0.21$ Gyrs (He
atmosphere result).  Objects six and seven are much more reliable than
objects nine or even eight, however.  For the most likely situation of
Hydrogen atmospheres, these objects are $0.98 \pm 0.04$ Gyrs, and $1.14
\pm 0.06$ Gyrs.  The cluster distance uncertainty of $\pm 0.08$ increases
these age uncertainties slightly, to $\pm 0.06$ and $\pm 0.09$ Gyrs,
respectively.  The WD cooling ages are not surprising since the accepted
age for NGC 188 is $\approx 6$ Gyrs.  Observations probing to V = $27$ or
$28$ will be necessary to find the faintest cluster WDs.  These faintest
cluster WDs will be both redder and superimposed on yet more background
objects, however, limiting the efficacy of the simple photometric
technique employed here.  A third filter sensitive to the Calcium H \& K
lines in the background stars may help, though deep observations with such
a narrow-band filter will be time-consuming.  Proper motions may also
help, though the proper motion difference between NGC 188 and the Galactic
field is only $\sim1$ mas/yr (Mendez \& van Altena 1996), requiring a ten
or twenty year delay before a meaningful second epoch could be obtained.
Perhaps the best means of obtaining the WD sequence terminus would be to
observe essentially the entire cluster.  NGC 188 is expected to have $\geq
6$\% by mass of its stars in the form of WDs (von Hippel 1998).  For a
cluster dynamically evolved from an IMF with a Salpeter-like slope, the
number fraction of WDs should be at least as great as the mass fraction.
The proper motion study of Dinescu \etal (1996) found $360$ cluster
members to B = $16.2$, so it seems reasonable to assume that at least
$1000$ cluster members exist (see, for comparison, the V-band LF in Figure
8a).  Since more than $50$\% of cooling WDs pile up near the terminus of
the WD sequence, we expect $\geq 30$ WDs to define this cooling sequence
terminus, and they should thus be observable as an excess even against a
significant background contamination.

Although we have not been able to find the faintest WDs in NGC 188, our
result demonstrates the value of the WD age technique and its very low
internal errors.  While observationally difficult, the theoretical
age determination based on WD luminosities, once a good cluster WD sample
has been isolated, is much less troublesome than the theoretical age
determination based on isochrone fitting to the main sequence turn-off.

\section{Conclusion}

We have employed precise and carefully calibrated V- and I-band photometry
of NGC 188 at WIYN to explore the cluster luminosity function and study
the cluster white dwarfs.  Our observations span more than a year and
include a single night tied to the Landolt (1992) Johnson-Kron-Cousins
photometric system to within $0.01$ mag.  Photometric bootstrapping
procedures were employed to calibrate the faint stars from the remainder
of the nights to the single calibrated night and the total external
systematic in the faint star calibration should be $\leq 0.032$ mag.  Our
photometry is offset by V = $0.052$ (fainter) from Sandage (1962) and
Eggen \& Sandage (1969).  Since all photometry for the past three decades
have been tied to Sandage (1962) and Eggen \& Sandage (1969), all past
photometry includes a $0.05 \pm 0.01$ mag photometric zero point error.

We employ the Pinsonneault \etal (1998) fiducial open cluster main
sequence to derive distance moduli to NGC 188 based on the two most
frequently used cluster reddening values, E(B$-$V) = $0.09$ and $0.12$.
Our best fit distance moduli are m$-$M = $11.43$ for E(B$-$V) = $0.09$ and
m$-$M = $11.67$ for E(B$-$V) = $0.12$.  Based on past distance estimates
to NGC 188 (see Twarog \etal 1997) we favor the E(B$-$V) = $0.09$ result.
Carefully considering a variety of sources of external error, we estimate
that the NGC 188 distance modulus is $11.43 \pm 0.08$ and E(B$-$V) =
$0.09$, with the largest single source of uncertainty being the uncertainty 
in the metallicity of the cluster.

In order to determine the cluster luminosity function we employ the Galaxy
model of Reid \& Majewski (1993) to determine the number of contaminating
field stars.  We find this technique works well.  We also find that our
morphological ($0.7${\arcsec} seeing) and color selections decrease the
expected extragalactic contamination to less than one object in either the
white dwarf or main sequence regions of the CMD.  We employ the standard
approach of artificial star experiments to estimate completeness and find
our observations are $\geq 50$\% complete along the main sequence to V =
$24.6$.  After determining the small field star and completeness
corrections we find that the NGC 188 central-field LF peaks at 
M$_I \approx 3$ to $4$.
This is unlike the solar neighborhood LF and unlike the LFs of dynamically
unevolved portions of open and globular clusters, all of which typically
rise continuously until M$_I \approx 9.5$.  Although we find that $\geq
50$\% of the unresolved objects in this cluster are multiple systems with
mass ratios $\geq 0.3$, their presence cannot account for the shape of the
NGC 188 LF.  For theoretical reasons (Terlevich 1987; Vesperini \& Heggie
1997) having to do with the long-term survivability of NGC 188 (McClure \&
Twarog 1977; Binney \& Tremaine 1987) we believe the cluster had a typical
IMF, but that it is now highly
dynamically evolved and the missing low luminosity stars 
are either in the cluster outskirts, or have left
the cluster altogether.

We identify nine candidate WDs in NGC 188, of which we expect at least
three, and perhaps six, are bona fide cluster WDs.  The luminosities of
the faintest likely WD indicates an age (Bergeron \etal 1995) of $1.14 \pm
0.09$ Gyrs, where the error in age includes the cluster distance
uncertainty and we assume the WD has a Hydrogen atmosphere.  This age is a
lower limit to the cluster age and observations probing to V = $27$ or
$28$ will be necessary to find the faintest cluster WDs and independently
determine the cluster age.  While our lower age limit is not surprising,
since the accepted cluster age is $\approx 6$ Gyrs, our result
demonstrates the value of the WD age technique with its very low internal
errors.  While observationally difficult, the theoretical age
determination based on WD luminosities, once a good cluster WD sample has
been isolated, is much less troublesome than the theoretical age
determination based on isochrone fitting to the main sequence turn-off.

\acknowledgments

We thank Betsy Green for many useful discussions and help with calibration
issues.  We also thank Con Deliyannis, Pierre Demarque, Bob Mathieu, 
Imants Platais, and an anonymous referee for helpful comments on the
manuscript.  Ted von Hippel expresses his appreciation for grant support
provided by the Edgar P. and Nona B.  McKinney Charitable Trust.  Ata
Sarajedini expresses his gratitude to UCO/Lick Observatory for kind
hospitality during his visit.  Ata Sarajedini was supported by the
National Aeronautics and Space Administration (NASA) grant number
HF-01077.01-94A from the Space Telescope Science Institute, which is
operated by the Association of Universities for Research in Astronomy,
Inc., under NASA contract NAS5-26555.


\newpage

\figcaption[fig1.ps]
{A. The correlation between the universal time of observation and the mean
V-band photometric residual after removing the first-order extinction and
color terms.  The error bars on the NGC 188 data points (``o'' symbols) represent
the standard deviation of at least $16$ stars observed at that epoch.  The
mean value for the standards at each epoch are represented by ``x'' symbols.  The
absolute difference in the standards and the relative difference in the
NGC 188 data both yield essentially the same slope.  The dashed line shows
the slope fit only to the standards whereas the solid line shows the slope
fit to the NGC 188 data and imposed on the standard star data.  The zero
point is set at UT = $10.20$.  B. Same as A for the I-band.}

\figcaption[fig2.ps]
{A. The difference between the V-band photometry of Eggen \& Sandage
(1969) and our V-band photometry versus apparent magnitude.  Dashed lines
at $\Delta$ V = $0.0$ and $-0.1$ are overplotted to aid the eye in
comparing the photometry.  B.  Same as A but plotted against V$-$I color.}

\figcaption[fig3.ps]
{The SExtractor (Bertin \& Arnouts 1996) ``Stellarity Index''
classifications versus I-band magnitude.  The Stellarity index ranges from
$1.0$ for definite stars to $0.0$ for definite galaxies.  The horizontal
dashed line at $0.95$ is our high-confidence star threshold.  As expected,
classification breaks down for the faintest objects.}

\figcaption[fig4.ps]
{The calibrated color-magnitude diagram in V and I for NGC 188.  Typical
photometric errors at each integer V-band magnitude starting at V = $19$
are presented down the right-hand side.  Only objects with a likelihood
$\gtrsim 95$\% of being stars according to SExtractor are plotted.  The
loci of solar neighborhood white dwarfs in the trigonometric data of Monet
\etal (1992), offset to the cluster distance and reddening, is overplotted
as a dashed line.}

\figcaption[fig5.ps]
{A. Our photometry in the region of the Pinsonneault \etal (1998) fiducial
solar metallicity main sequence (solid line) for our best fit distance of
m$-$M = $11.43$ for E(B$-$V) = $0.09$.  The one sigma error bars are
plotted for the data and fiducial sequences offset by $\pm 0.05$ mags are
presented as dashed lines.  The objects used in the distance fit are
identified with circles.  B. Similar to A but for our best fit distance of
m$-$M = $11.67$ for E(B$-$V) = $0.12$.}

\figcaption[fig6.ps]
{A. The distribution of our selected stars about the Pinsonneault \etal
(1998) fiducial main sequence for an {\it initial} cluster distance and reddening of
m$-$M = $11.41$ and E(B$-$V) = $0.09$.  The effect of binary star systems
is clearly present in the tail of the distribution towards positive offset
values.  The first three or four bins contain single stars or q $\leq 0.3$
binaries, while the other bins contain q $> 0.3$ binary systems.  B.
Similar to A for {\it initial} cluster values of m$-$M = $11.63$ and
E(B$-$V) = $0.12$.}

\figcaption[fig7.ps]
{The color-magnitude diagram in V and I from the Reid \& Majewski 
Galaxy model for the
field location and size NGC 188.  This model population serves as our
field star sample.}

\figcaption[fig8.ps]
{A. The upper histogram marked ``data'' is the V-band luminosity function
of observed cluster and field stars selected to have colors and magnitudes
near the fiducial main sequence of NGC 188.  The lower histogram marked
``model'' is the V-band luminosity function of model field stars with the
same colors and magnitudes from the Reid \& Majewski Galaxy model.  The completeness
fraction is shown as the solid line overplotting the ``+'' symbols, dropping
from $1.0$ near V = $20$ to $0.0$ near V = $25$.  B. Similar to A
but for the I-band.  C. The absolute magnitude V-band main sequence
luminosity function derived from the completeness-corrected difference between 
the observed CMD
counts and the Reid \& Majewski model CMD counts and based on the cluster 
distance and
reddening.  D. Similar to C but for the I-band.  In addition, a number of
other LFs are over-plotted:  the solid line with the ``+'' symbols for the
solar neighborhood, the solid lines with the ``o'' symbols for NGC 2477,
the solid lines with the ``*'' symbols for NGC 2420, the dotted line for
NGC 6397, the dashed line for 47 Tuc, the dash single-dotted line for M15,
and the dash double-dotted line for $\omega$ Cen.  Some of the globular
cluster LFs are on the $M_{814}$ (HST I-band magnitude) system, but this
is very close to the standard Cousins I-band magnitude system.  The
cluster LFs have been scaled to fit on the same plot.}

\figcaption[fig9.ps]
{The de-reddened color-magnitude diagram of Figure 4 with distance
removed.  The nine candidate WDs (two overlap) are identified by overplotted 
circles.  In addition to the observed Monet \etal (1992) WD sequence, the
log(g) = $8$ Hydrogen atmosphere theoretical WD cooling sequence of
Bergeron \etal (1995) is plotted (as the solid line).}

\clearpage

\begin{deluxetable}{rcllcl}
\tablewidth{0pt}
\tablecaption{Log of Observations}
\tablehead{
\colhead{date} & \colhead{conditions} & \colhead{V} & \colhead{I} &
\colhead{seeing}  & \colhead{notes} \nl
(1)\phantom{1234567} & (2) & (3) & (4) & (5) & \phantom{1234567890}(6)}
\startdata
Jul. 10, 1995    & NP & \nodata & 1.0  & 1.0-1.2 & \nodata                  \nl
Jul. 24-27, 1995 & P  & 4.25    & 2.5  & 1.0-1.5 & standards but non-linear \nl
Feb. 4, 1996     & P  & \nodata & 2.0  & 0.7     & no standards             \nl
Aug. 12-15, 1996 & NP & 2.5     & 3.75 & 1.2-2.0 & discarded                \nl
Sep. 19, 1996    & P  & 1.5     & 1.0  & 0.9-1.6 & standards and post-repair
\enddata
\end{deluxetable}


\begin{deluxetable}{crcrcrcc}
\tablewidth{0pt}
\tablecaption{Parameters for Candidate White Dwarfs}
\tablehead{
\colhead{ID} & \colhead{(V$-$I)$_{\rm o}$} & \colhead{M$_{\rm V}$} & 
\colhead{T$_{\rm eff}$} & \colhead{Mass} & \colhead{M$_{\rm bol}$} & 
\colhead{log(Age)} & \colhead{model} \nl
(1) & (2)\phantom{12} & (3) & (4)\phantom{1234} & (5) & (6)\phantom{1234} & (7) & (8)}
\startdata
1       &  0.03   & 11.22 (0.035) & 15115 (301) & 0.614 & 10.00 (0.088) & 8.339 (0.025) & H  \nl
\nodata & \nodata & \nodata       & 15336 (234) & 0.588 &  9.99 (0.065) & 8.376 (0.023) & He \nl
2       & $-$0.31 & 11.92 (0.043) & 10720 (152) & 0.604 & 11.51 (0.062) & 8.735 (0.016) & H  \nl
\nodata & \nodata & \nodata       & 11443 (196) & 0.580 & 11.28 (0.075) & 8.768 (0.021) & He \nl
3       & $-$0.04 & 12.06 (0.044) & 10265 (132) & 0.604 & 11.71 (0.056) & 8.782 (0.014) & H  \nl
\nodata & \nodata & \nodata       & 10835 (169) & 0.579 & 11.52 (0.070) & 8.834 (0.019) & He \nl
4       & $-$0.03 & 12.08 (0.044) & 10212 (132) & 0.603 & 11.73 (0.057) & 8.788 (0.014) & H  \nl
\nodata & \nodata & \nodata       & 10768 (168) & 0.579 & 11.54 (0.070) & 8.841 (0.019) & He \nl
5       &  0.37   & 12.24 (0.046) &  9763 (120) & 0.602 & 11.93 (0.054) & 8.837 (0.013) & H  \nl
\nodata & \nodata & \nodata       & 10149 (174) & 0.579 & 11.80 (0.073) & 8.910 (0.019) & He \nl
6       &  0.33   & 12.78 (0.061) &  8495 (132) & 0.598 & 12.54 (0.069) & 8.989 (0.017) & H  \nl
\nodata & \nodata & \nodata       &  8486 (158) & 0.575 & 12.58 (0.081) & 9.108 (0.020) & He \nl
7       &  0.24   & 13.01 (0.076) &  8003 (156) & 0.597 & 12.80 (0.085) & 9.055 (0.022) & H  \nl
\nodata & \nodata & \nodata       &  7923 (172) & 0.575 & 12.88 (0.095) & 9.184 (0.024) & He \nl
8       &  0.63   & 13.04 (0.078) &  7943 (158) & 0.597 & 12.83 (0.088) & 9.064 (0.022) & H  \nl
\nodata & \nodata & \nodata       &  7857 (172) & 0.575 & 12.92 (0.095) & 9.193 (0.024) & He \nl
9       &  0.72   & 13.83 (0.147) &  6516 (225) & 0.592 & 13.70 (0.152) & 9.280 (0.037) & H  \nl
\nodata & \nodata & \nodata       &  6407 (220) & 0.574 & 13.81 (0.150) & 9.411 (0.035) & He
\enddata
\end{deluxetable}


\begin{references}

\reference{}
Baliunas, S. L., \& Guinan, E. F. 1985, \apj, 294, 207

\reference{}
Bergeron, P., Wesemael, F., \& Beauchamp, A. 1995, \pasp, 107, 1047

\reference{}
Bertin, E., \& Arnouts, S. 1996, \aaps, 117, 393

\reference{}
Binney, J., \& Tremaine, S. 1987, in Galactic Dynamics (Princeton
University Press, Princeton, NJ), p. 491

\reference{}
Caputo, F., Chieffi, A., Castellani, V., Collados, M., Martinez Roger, C.,
\& Paez, E. 1990, \aj, 99, 261

\reference{}
Cardelli, J. A., Clayton, G. C., \& Mathis, J. S. 1989, \apj, 345, 245

\reference{}
Carraro, G., \& Chiosi, C. 1994, \aap, 288, 751

\reference{}
Carraro, G., \& Chiosi, C., Bressan, A., \& Bertelli, G. 1994, \aap, 103, 375

\reference{} 
De Marchi, G., \& Paresce, F. 1995a, \aap, 304, 202

\reference{} 
De Marchi, G., \& Paresce, F. 1995b, \aap, 304, 211

\reference{}
Demarque, P., Guenther, D. B., \& Green, E. M. 1992, \aj, 103, 151

\reference{}
Dinescu, D. I., Demarque, P., Guenther, D. B., \& Pinsonneault, M. H. 1995,
\aj, 109, 2090

\reference{}
Dinescu, D. I., Girard, T. M., van Altena, W. F., Yang, T. -G., Lee, Y. -W.
1996, \aj, 111, 1205

\reference{}
Eggen, O. J., \& Sandage, A. R. 1969, \apj, 158, 669

\reference{} 
Elson, R. A. W., Gilmore, G. F., Santiago, B. X., \& Casertano, S. 1995, \apj,
110, 682

\reference{}
ESA 1997, The Hipparcos and Tycho Catalogues, ESA SP-1200

\reference{}
Gilmore, G., Reid, I. M., \& Hewitt, P. C. 1985, \mnras, 213, 257

\reference{}
Gilmore, G., Wyse, R. F. G., \& Kuijken, K. 1989, \araa, 27, 555

\reference{}
Harris, W. E., Fitzgerald, M. P., \& Reed, B. C.  1981, \pasp, 93, 507

\reference{}
Henry, T. J., \& McCarthy, D. W. 1993, \aj, 106, 773

\reference{}
Hobbs, L. M., Thorburn, J. A., \& Rodriguez-Bell, T. 1990, \aj, 100, 710

\reference{}
Hobbs, L. M., \& Pilachowski, C. 1988, \apj, 334, 734 

\reference{}
Kaluzny, J. 1990, AcA, 40, 61

\reference{}
Kaluzny, J., \& Shara, M. M. 1987, \apj, 314, 585

\reference{}
Keenan, D. W., Innanen, K. A., \& House, F. C. 1973, \aj, 78, 173

\reference{}
Keeping, E. S. 1962, Introduction to Statistical Inference (Princeton, Van
Nostrand), p. 202

\reference{} 
Kroupa, P., Tout C. A., \& Gilmore, G. 1993, \mnras, 262, 545

\reference{}
Landolt, A. U. 1983, \aj, 88, 439

\reference{}
Landolt, A. U. 1992, \aj, 104, 340

\reference{}
Leonard, P. J. T., \& Linnell, A. P. 1992, \aj, 103, 1928

\reference{}
Massey, P., Jacoby, G., Carder, E., \& Harris, H. 1987, NOAO Newsletter,
12, 28

\reference{}
Mathieu, R. D., \& Dolan, C. 1998, in preparation

\reference{}
McClure, R. D., \& Twarog, B. A. 1977, \apj, 214, 111

\reference{}
Mendez, R. A., \& van Altena, W. F. 1996, \aj, 112, 655

\reference{}
Monet, D. G., Dahn, C. C., Vrba, F. J., Harris, H. C., Pier, J. R., Luginbuhl,
C. B., \& Ables, H. D. 1992, \aj, 103, 638

\reference{}
Moss, D. 1985, \aap, 150, 343

\reference{}
Norris, J., \& Smith, G. H. 1985, \aj, 90, 2526

\reference{} 
Paresce, F., De Marchi, G., \& Romaniello, M. 1995, \apj, 440, 216

\reference{}
Phelps, R., Janes, K. A., \& Montgomery, K. A. 1994, \aj, 107, 1079

\reference{}
Pinsonneault, M. H., Stauffer, J., Soderblom, D. R., King, J. R., \& Hanson,
R. B. 1998, \apj, in press 

\reference{}
Pols, O. R., \& Marinus, M. 1994, \aap, 288, 475

\reference{}
Reid, I. N., \& Majewski, S. R. 1993, \apj, 409, 635

\reference{}
Reid, I. N., Yan, L., Majewski, S., Thompson, I., \& Smail, I. 1996, \aj,
112, 1472

\reference{}
Richer, H. B., \etal 1997, \apj, 484, 741

\reference{}
Sandage, A. 1962, \apj, 135, 333


\reference{}
Scott, J. E., Friel, E. D., \& Janes, K. A. 1995, \aj, 109, 1706

\reference{}
Smail, I., Hogg, D. W., Yan, L., \& Cohen, J. G. 1995, \apj, 449, L105

\reference{}
Spinrad, H., Greenstein, J. L., Taylor, B. J., \& King, I. R. 1970, \apj,
162, 891

\reference{}
Spinrad, H., \& Taylor, B. J. 1969, \apj, 157, 1279

\reference{}
Stetson, P. B. 1994, \pasp, 106, 250

\reference{}
Terlevich, E. 1987, \mnras, 224, 193

\reference{}
Tody, D. 1993, in Astronomical Data Analysis Software and Systems II,
A.S.P. Conference Ser., Vol 52, eds. R. J. Hanisch, R. J. V. Brissenden, \&
J. Barnes, p. 173

\reference{}
Twarog, B. A. 1978, \apj, 220, 890

\reference{}
Twarog, B. A., \& Anthony-Twarog, B. J. 1989, \aj, 97, 759

\reference{}
Twarog, B. A., Ashman, K. M., \& Anthony-Twarog, B. J. 1997, \aj, 114, 2556

\reference{}
Upgren, A. R., Mesrobian, W. S., \& Keridge, S. J. 1972, \aj, 77, 74

\reference{}
van 't Veer, F. 1984, \aap, 139, 477

\reference{}
Vesperini, E., \& Heggie, D. C. 1997, \mnras, 289, 898

\reference{}
von Hippel, T. 1998, \aj, 115, 1536

\reference{}
von Hippel, T., Gilmore, G., \& Jones, D. H. P. 1995, \mnras, 273, L39

\reference{}
von Hippel, T., Gilmore, G., Tanvir, N., Robinson, D., \& Jones, D. H. P.
1996, \aj, 112, 192

\end{references}
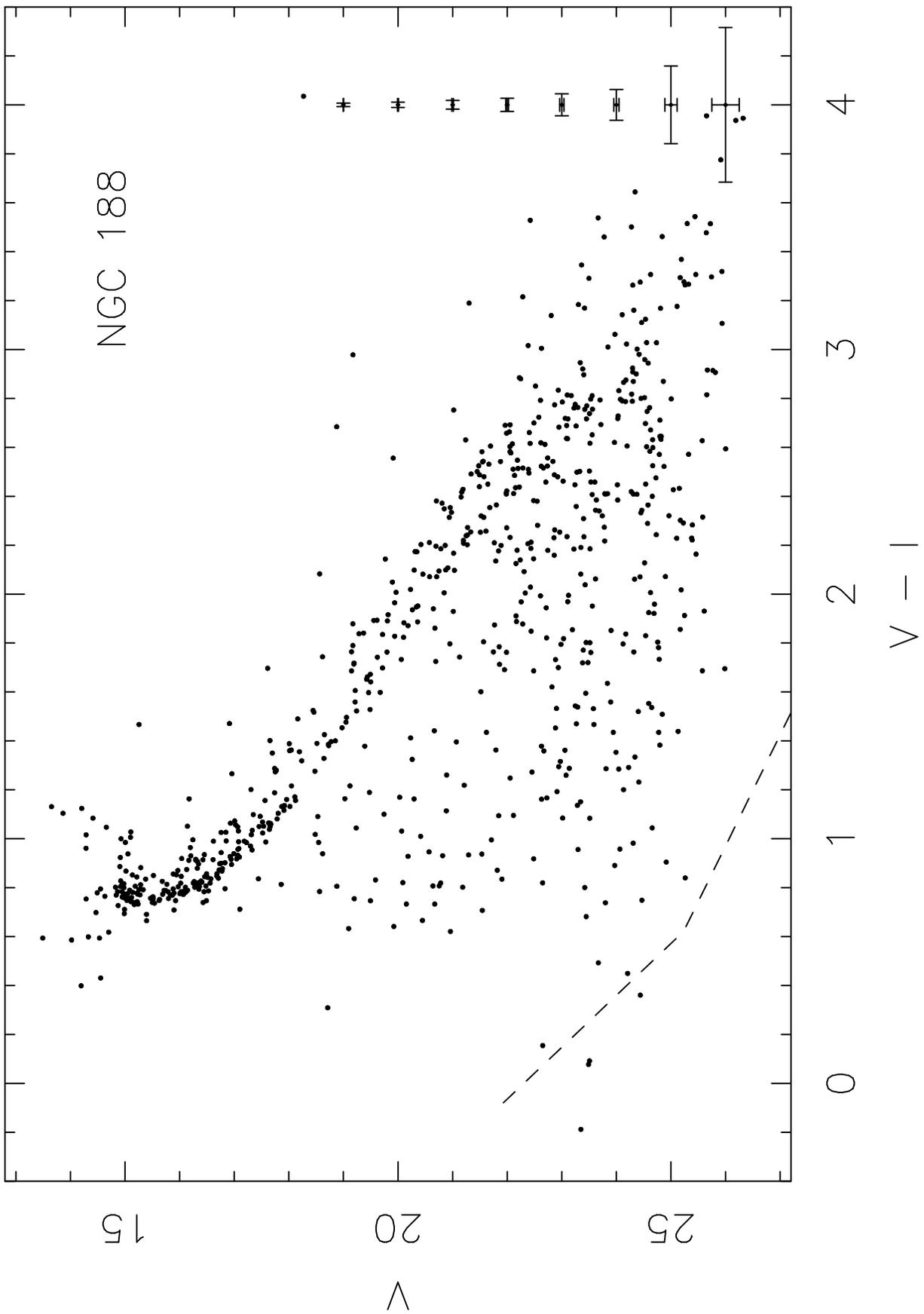
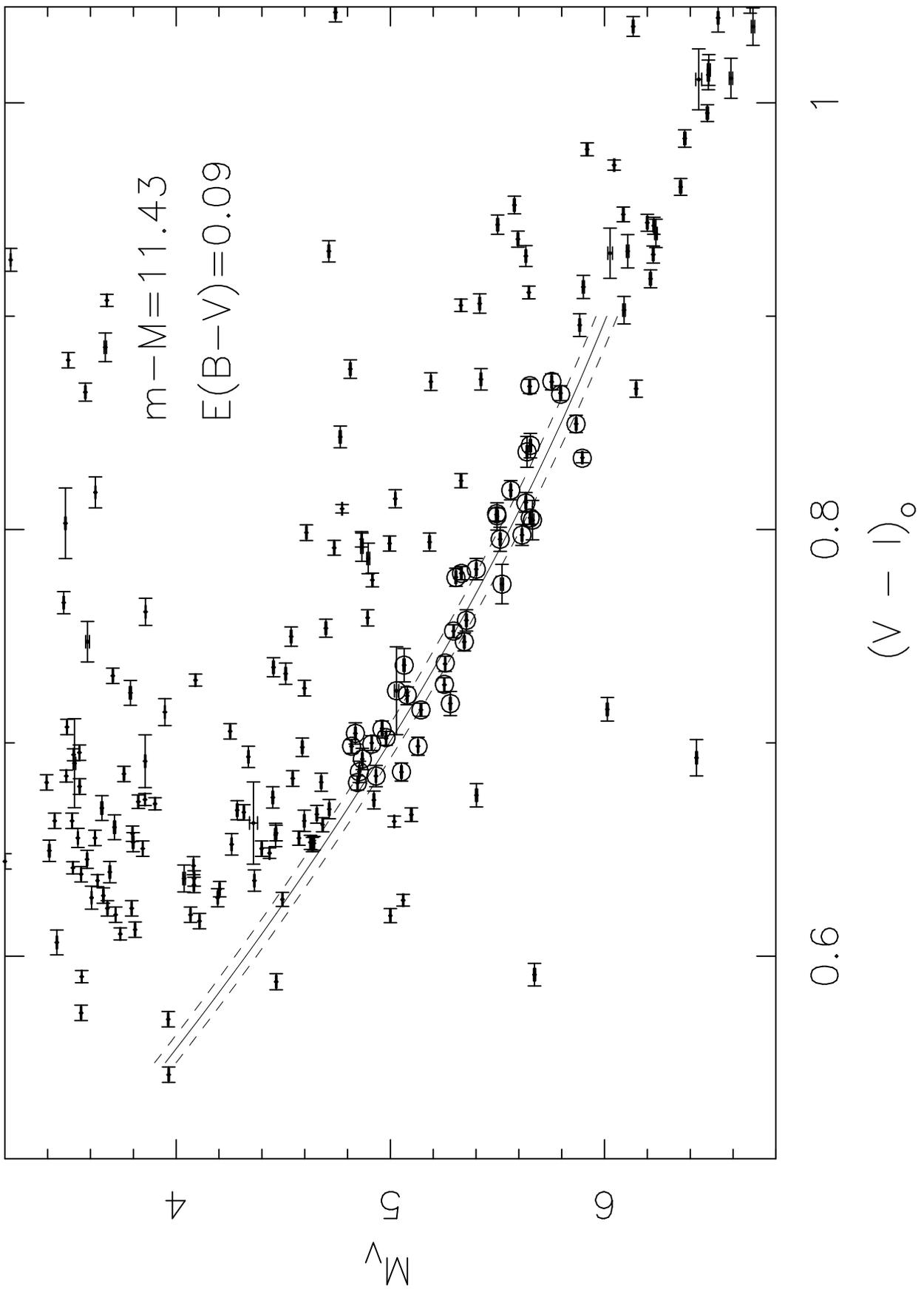
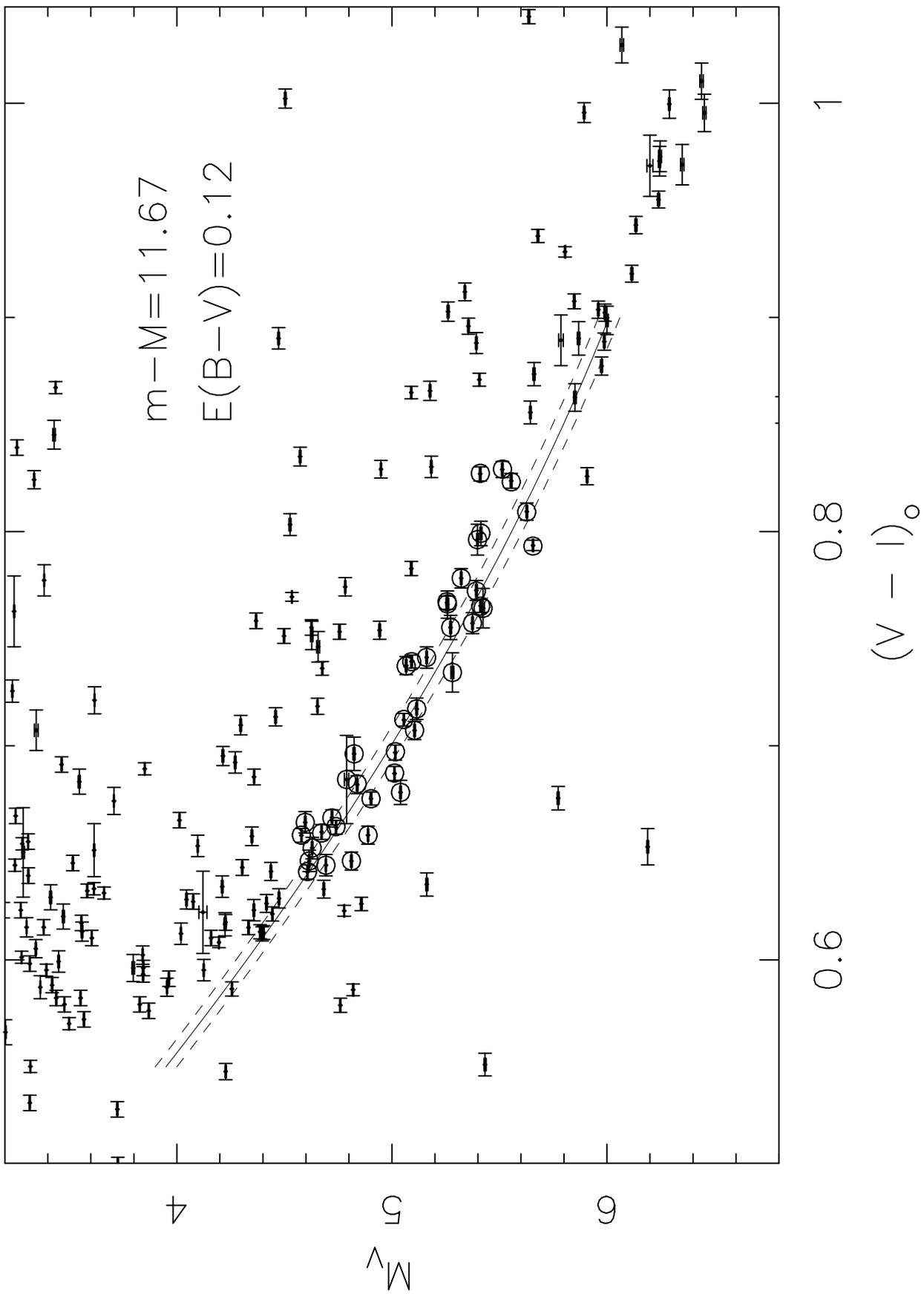
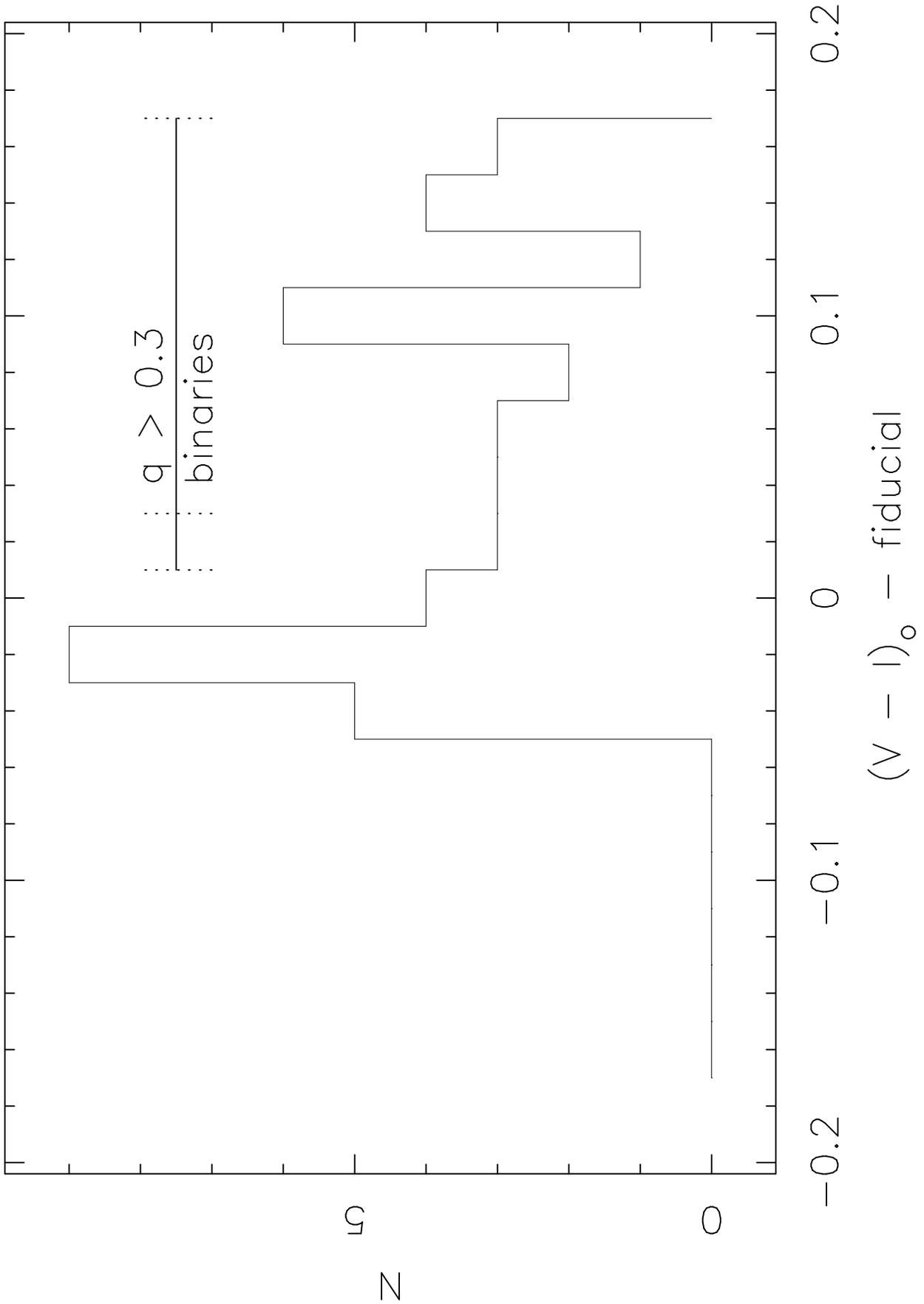
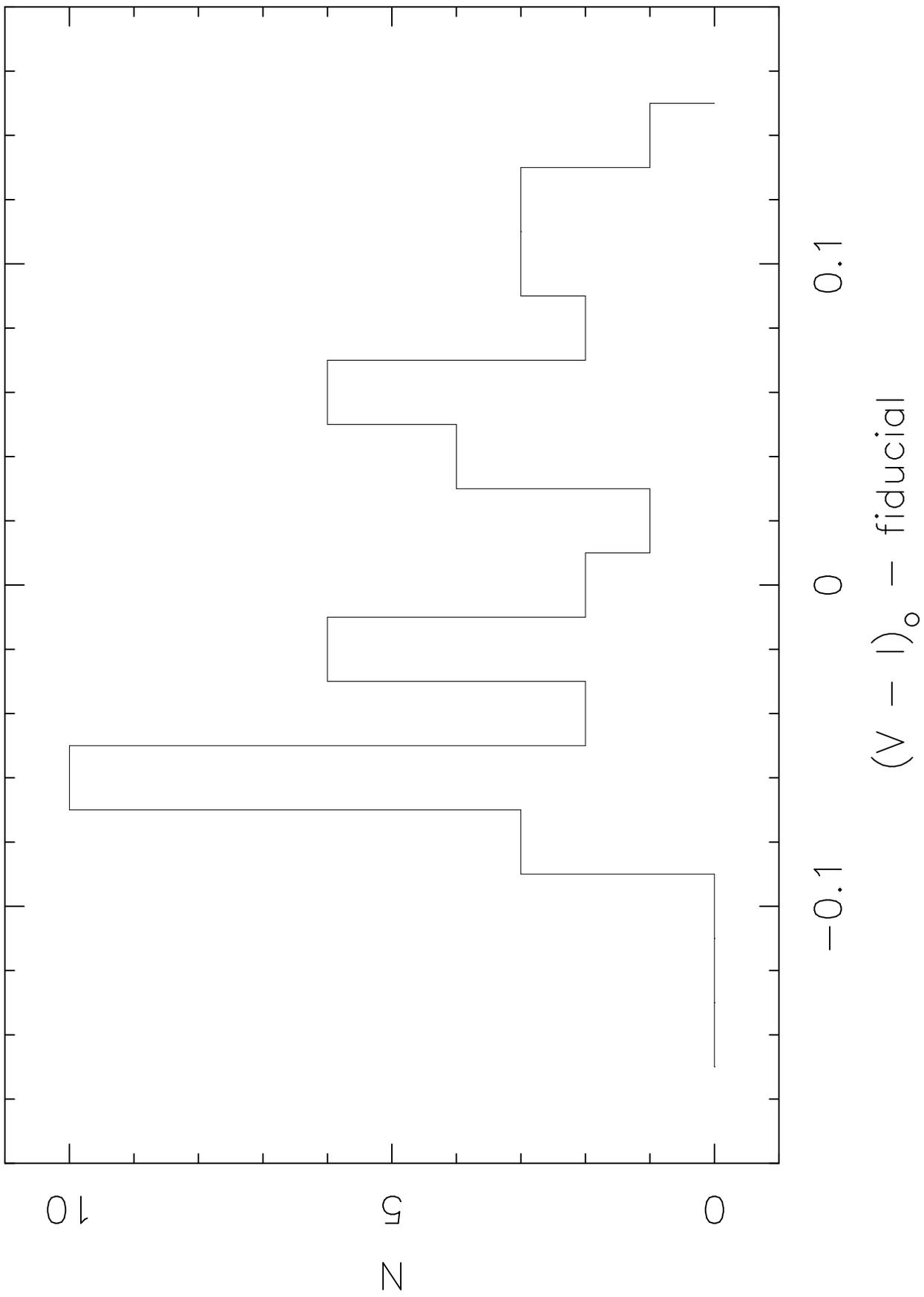
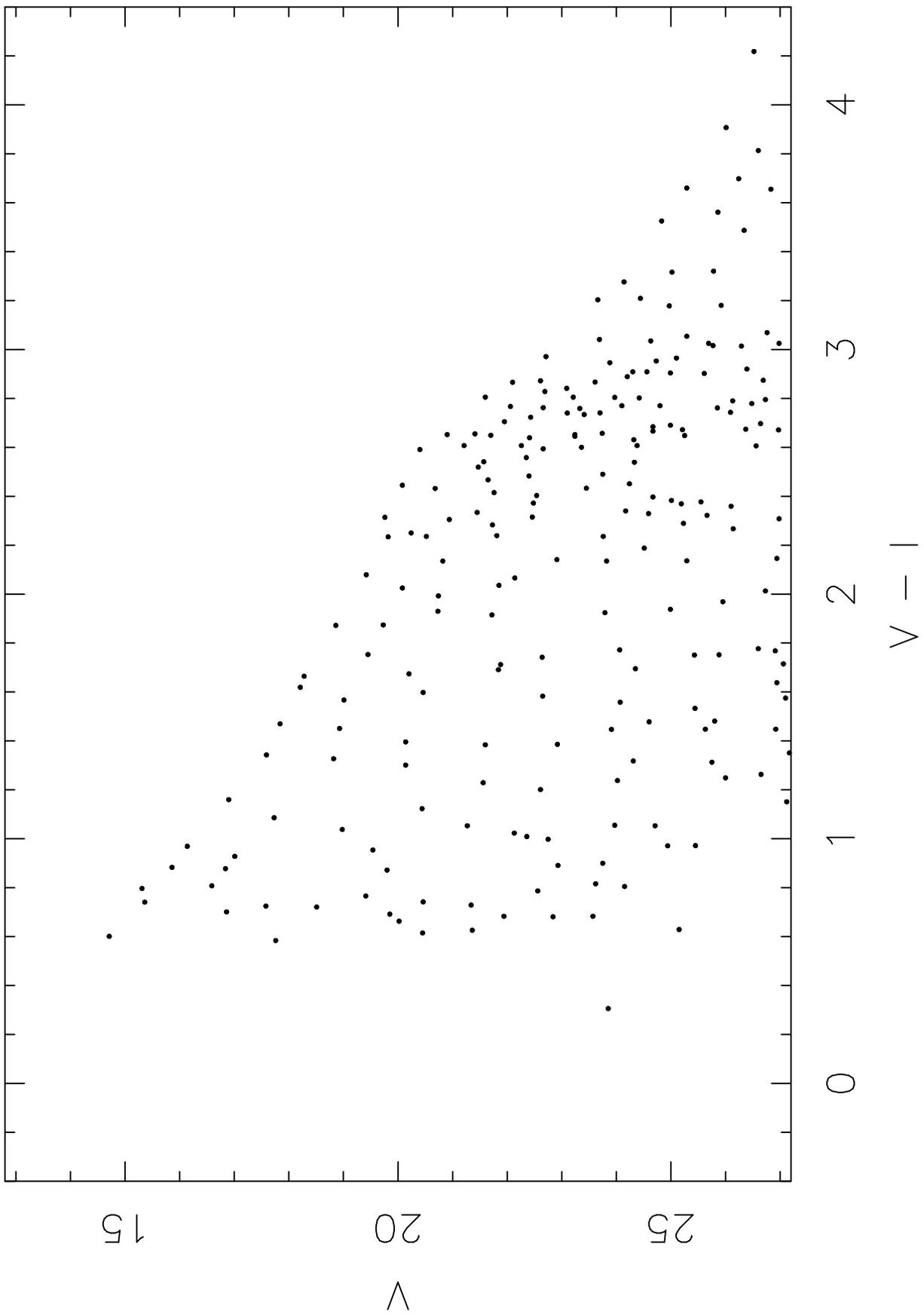
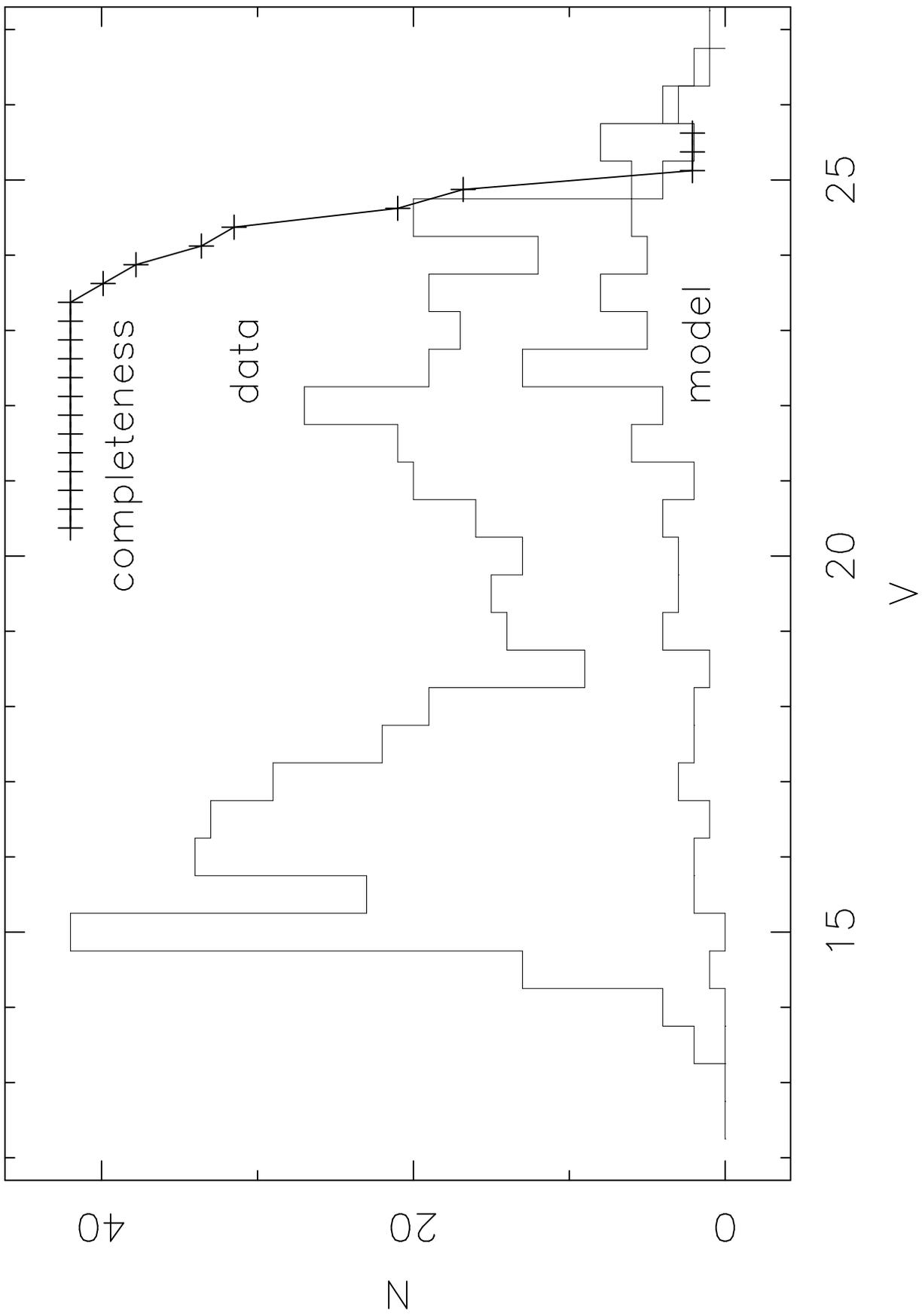
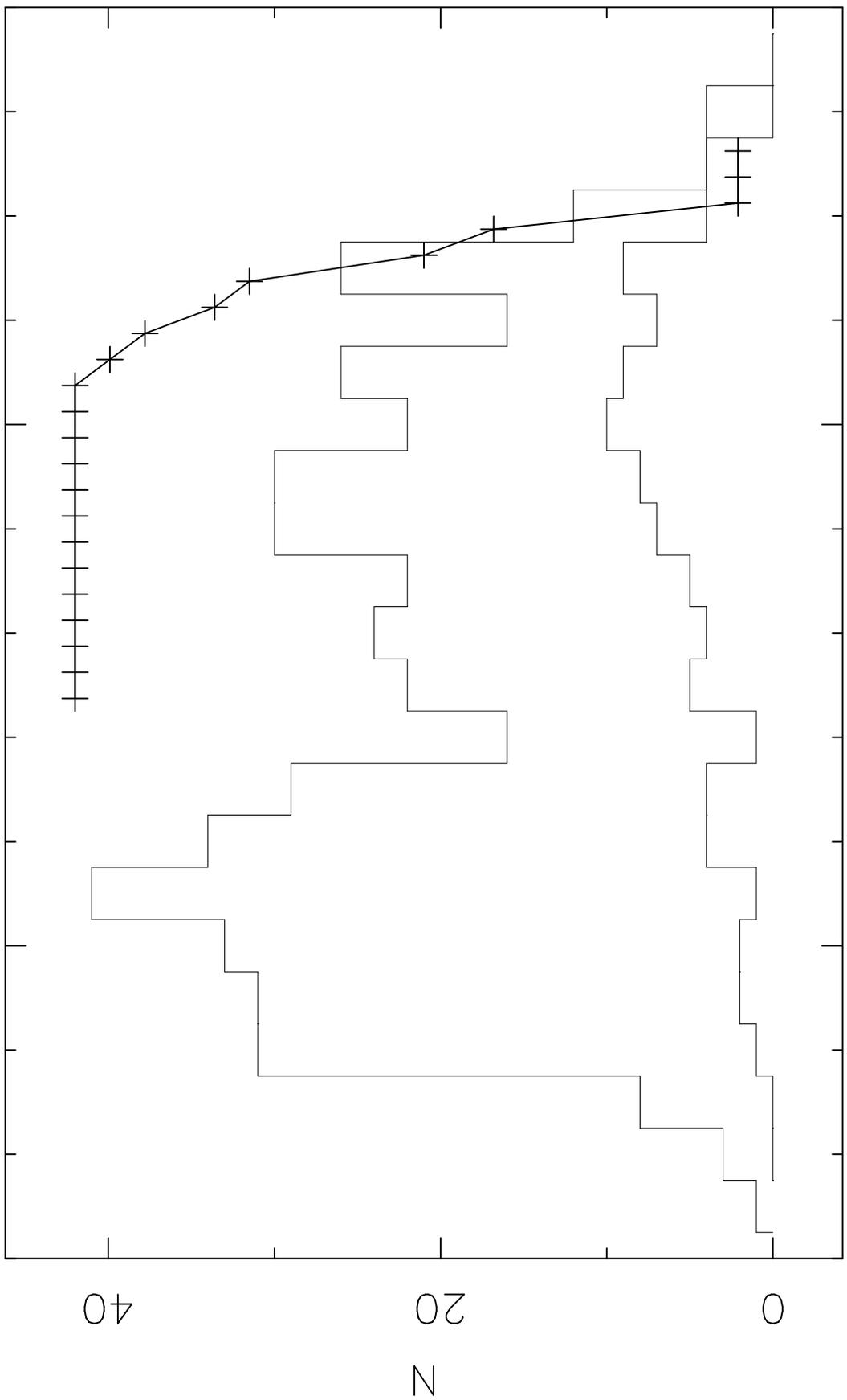
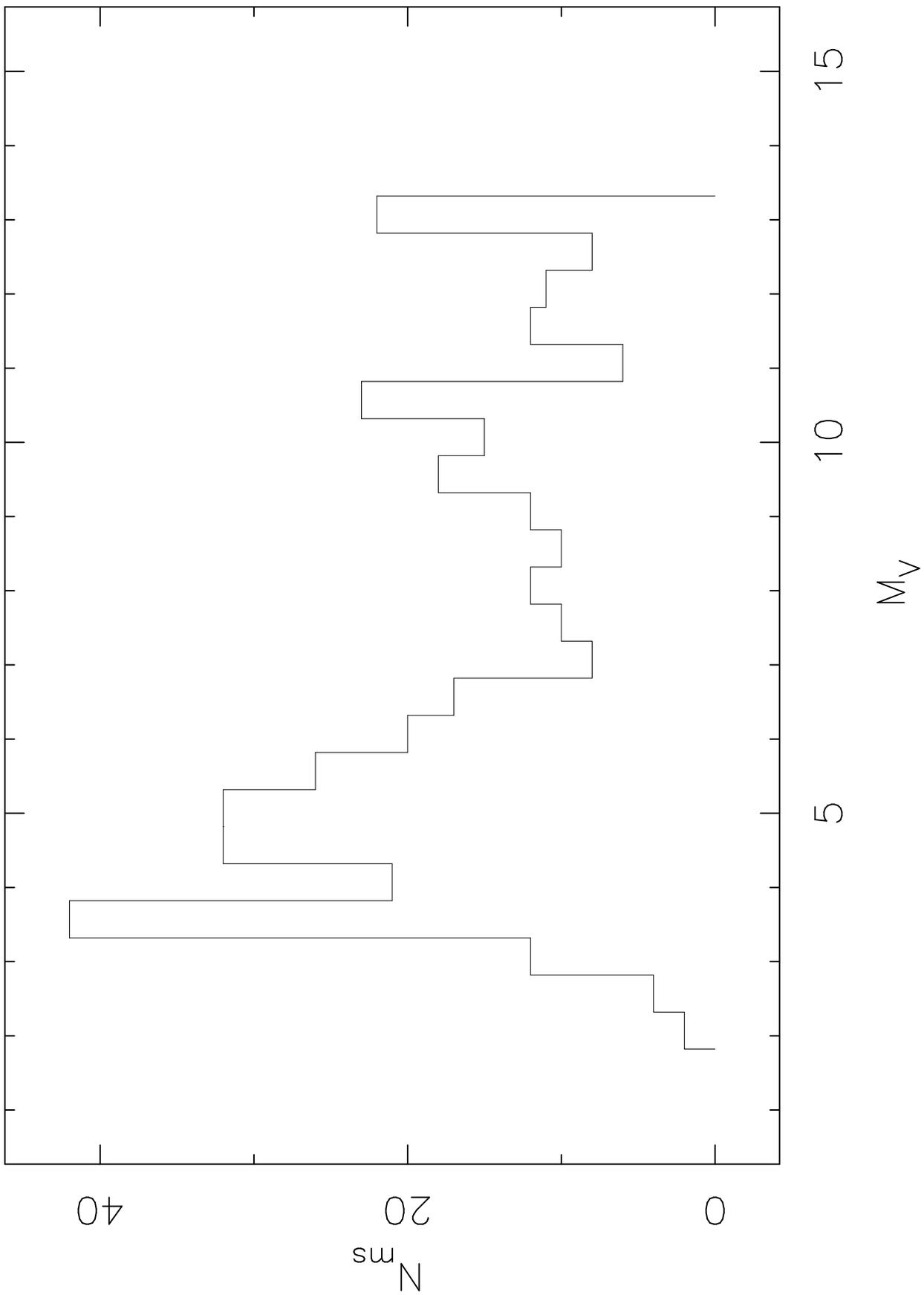
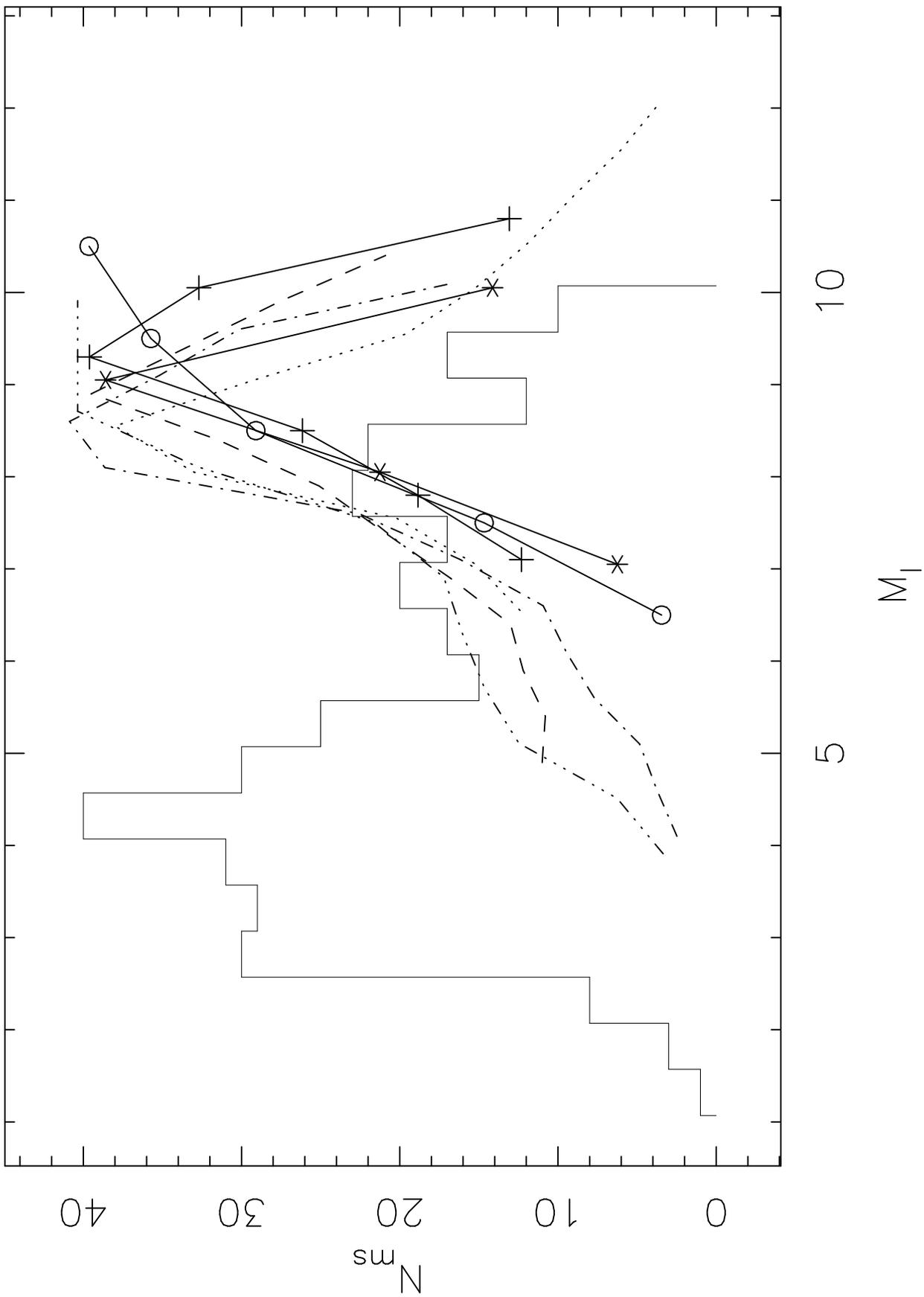
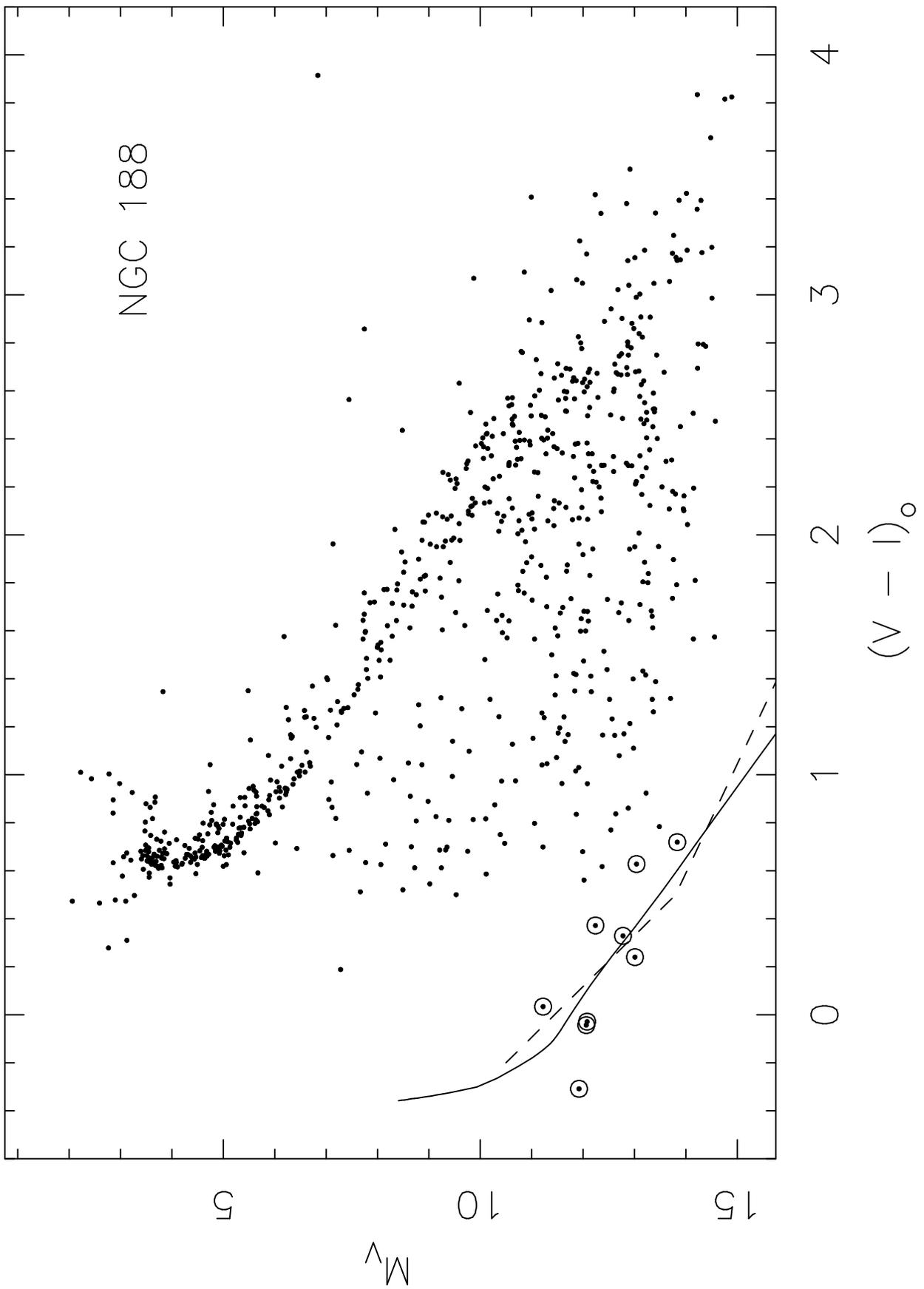
\end{document}